\documentclass[iicol]{sn-jnl}

\usepackage{graphicx}%
\usepackage{multirow}%
\usepackage{amsmath,amssymb,amsfonts}%
\usepackage{amsthm}%
\usepackage{mathrsfs}%
\usepackage[title]{appendix}%
\usepackage{xcolor}%
\usepackage{textcomp}%
\usepackage{manyfoot}%
\usepackage{booktabs}%
\usepackage{algorithm}%
\usepackage{algorithmicx}%
\usepackage{algpseudocode}%
\usepackage{listings}%
\usepackage{float}
\usepackage[authoryear]{natbib}

\theoremstyle{thmstyleone}%
\theoremstyle{thmstyletwo}%

\theoremstyle{thmstylethree}%

\chardef\us=`\_

\begin{document}

\title[Article Title]{Forbush Decreases during strong Geomagnetic Storms: Time Delays, Rigidity Effects, and ICME-Driven Modulation}

\author*[1,2]{\fnm{O.} \sur{Ahmed}}\email{osmanr18@ymailcom/osmanma24s@gmail.com}

\author[1]{\fnm{B.} \sur{Badruddin}}\email{badr.physamu@gmail.com}
\equalcont{These authors contributed equally to this work.}

\author[1]{\fnm{M.} \sur{Derouich}}\email{ derouichmoncef@gmail.com}
\equalcont{These authors contributed equally to this work.}

\affil*[1]{\orgdiv{Astronomy and Space Science Department}, \orgname{King Abdulaziz University}, \orgaddress{\street{Jeddah 21589}, \city{Jeddah}, \postcode{80203}, \state{Makkah}, \country{Saudi Arabia}}}

\affil[2]{\orgdiv{Department of Physics, Faculty of Natural and computational Sciences}, \orgname{Debre Tabor University}, \city{Debre Tabor}, \postcode{272}, \state{Amhara}, \country{Ethiopia}}



\abstract{We present the relationship between Forbush decreases (FDs) and associated geomagnetic storms, as well as their connections to interplanetary (IP) solar wind parameters, using high resolution minute data. FDs were classified into groups based on main phase decrease steps, and each group was analyzed using superposed epoch analysis. The results reveal that fast, turbulent, high-field sheath structures form before and pass during the onset of coronal mass ejection (CME) driven FDs, whereas corotating interaction region (CIR) driven events exhibit delayed amplification and more perturbed dynamics. Time lags between the onset of FDs and geomagnetic storms were calculated and discussed, providing insights crucial for space weather forecasting.
Correlation analyses between FD amplitude and peak values of various IP parameters were performed and discussed.
The relationship between FDs and geomagnetic storms was analyzed, revealing that for CME-driven events, FD amplitudes exhibit a stronger correlation with moderate and strong geomagnetic storms compared to extreme storms. The weaker correlation during extreme CME-driven storms may result from complex magnetospheric responses caused by successive events and prolonged southward interplanetary magnetic field Bz, unlike the more direct responses observed in moderate and strong single-event storms.
Interplanetary coronal mass ejection (ICME) manifestations were also correlated with FD amplitude, showing that events with fast forward shocks and compression sheath regions exhibit stronger correlations than those without shocks. Furthermore, we analyzed the energy dependence of FD amplitude using data from twelve neutron monitor stations at different latitudes and altitudes across the globe. As a result, the cosmic ray (CR) energy spectrum exhibits a two-step linear dependence with the FD amplitude, in the lower
rigidity FD amplitude decreases sharply, while in higher rigidity regimes, the decrease is more gradual.
A broader energy spectrum is recommended for more comprehensive conclusions.}




\maketitle

\section{Introduction} \label{intro}
Forbush decreases (FDs) are the short-term reduction in cosmic ray (CR) flux observed at Earth \citep{MISHEV20244160} often following high-speed solar wind streams \citep{melkumyan2018main} or coronal mass ejections (CMEs) \citep[e.g.,][]{2004ASSL..303.....D, 2009A&A...494.1107S}. CR intensity fluctuations are monitored by a global network of neutron monitors distributed across worldwide \citep[e.g.,][]{2004ApJ...601L.103B, 2011AdSpR..47.2210M, 2021JGRA..12628941V}.
These events typically coincide with geomagnetic storms, driven by interactions between solar wind structures and Earth's magnetosphere \citep[e.g.,][]{1987P&SS...35.1101G, 1994JGR....99.5771G, 2004JASTP..66.1121K}.
The correlation between FDs and magnetic storms is well-documented, particularly for CME-driven events \citep[e.g.,][]{2013ICRC...33.3583P}. Studies confirm that large FDs are predominantly caused by CMEs \citep[e.g.,][]{lingri2016forbush}.
The FD profile exhibits a notable dependence on magnetospheric conditions \citep[e.g.,][]{2023JASTP.25206146G}.
A notable characteristic of CR variation during geomagnetic disturbances is an increase in CR intensity \citep[e.g.,][]{1962JPSJS..17B.402K}.
Theoretical studies by \cite{2021RAA....21..234A} confirmed that CR modulation is predominantly influenced by geomagnetic indices (disturbance storm time [Dst], kp, and ap).
\cite{2023JASTP.24205981B} investigated the relationship between interplanetary (IP) parameters and the Dst index relation to CR intensity using data from eight neutron monitor stations. By applying wavelet analysis and correlation delay methods, they identified a strong link between CR count rate variations, solar wind speed, and Dst.\\
Although FDs and geomagnetic storms share a common underlying cause, the relationship between these phenomena, geomagnetic, and solar wind disturbances remains an area of ongoing interest \citep[e.g.,][]{YOSHIDA1966979, 2010AnGeo..28..479K, 2015SoPh..290..627C, 2015SoPh..290.1271B, 2020ApJ...896..133L, 2021RAA....21..234A}. While both may originate from solar and IP sources, their magnitudes do not exhibit a proportional relationship, likely due to different mechanisms producing the events \citep[e.g.,][]{1988JGR....93.2511Z, 1991SoPh..134..203B, 2008ASTRA...4...59A, 2009A&A...494.1107S, 2012ApJ...759..143A, MUSTAJAB201343, 2014SoPh..289.2669K}. Understanding this association is crucial for space weather forecasting and exploring the mechanisms behind CR modulation \citep[e.g.,][]{2010hesa.book.....S, 2023JASTP.24205981B}. Previous studies have highlighted the complex relationship between magnetic storms, CMEs, and FDs \citep[e.g.,][]{2013ICRC...33.3583P, 2019SpWea..17..487B}.
FD magnitudes correlate with the transient's orientation and the strength of the interplanetary magnetic field (IMF) within it, reaching their peak when complex solar wind structures reach the Earth \citep[e.g.,][]{2020BLPI...47...92R}.\\
Previous studies have employed statistical methods to explore the relationship between CME parameters and FDs. For instance, \cite{2016SoPh..291..285D} investigated the connection between FDs and CMEs associated with solar flares. Their findings indicated a correlation between FDs and various CME characteristics, such as angular width, source position, initial speed, and the class of the associated solar flare.
Recent studies have shown that FDs and geomagnetic storms do not exhibit a linear correlation.
\cite{2024Ge&Ae..64..289B} conducted a comprehensive analysis of IP parameters, FDs and their associated geomagnetic storms over a 65-year period (1957 -- 2022). By employing statistical methods, they investigated the relationship between these two phenomena and concluded that no linear correlation exists. Instead, they found that the connection between FDs and geomagnetic storms is influenced by the orientation of the IMF, particularly the south-north component Bz. Conversely, \cite{2024JASTP.26206305K} found that FD events are more strongly influenced by disturbances in the IMF B than by the Bz component.\\
Observational and theoretical studies have demonstrated the intricate relationships between solar wind parameters, geomagnetic activity, and CR flux modulation, emphasizing the role of solar wind-magnetosphere coupling in driving space weather phenomena \citep[e.g.][]{1988JGR....93.2511Z, 1998JGR...103.6917K, Ahmed2024a, Ahmed2024}. Theoretical models provide insights into how solar wind fluctuations influence magnetic storms and FDs, with variations in solar wind speed, IMF strength, and orientation significantly impacting FD amplitudes \citep[e.g.,][]{1999SSRv...89...21G, 2001ICRC....9.3507B, 2021AdSpR..68.4702B, 2023JASTP.24205981B}. Empirical evidence, supported by techniques such as superposed epoch analysis, has identified consistent patterns in plasma and IMF parameters across the heliosphere, as well as their effects on CR intensity \citep[e.g.,][]{2002SoPh..209..195B, 2021ApJ...910...99G}. These analyses have further explored the role of solar energetic particles \citep[e.g.,][]{2019SpWea..17.1765Y} and geomagnetic disturbances \citep[e.g.,][]{Ahmed2024a, Ahmed2024}, enhancing our understanding of modulation mechanisms that drive FDs. FD intensity has been linked to the product of plasma speed and IMF magnitude, while CME characteristics, such as stronger shock fronts and faster magnetic fields, are shown to enhance CR modulation \citep[e.g.,][and references therein]{2014SoPh..289.3949B, 2021AdSpR..68.4702B}.\\
In addition to superposed epoch analysis, correlation analysis has been employed to examine the relationship between FDs and IP parameters \citep[e.g.,][]{1977P&SS...25..681P, 2023AdSpR..71.2006S}.
Superposed epoch analysis \citep{1998P&SS...46.1015B, 2016Ap&SS.361..253B} effectively illustrates the time variation and lag of FD with respect to IP parameters, while correlation analysis quantifies the strength and nature of their relationship.
FD intensity is primarily governed by derived functions, typically represented as the product of solar wind plasma speed and IMF strength, effectively capturing their relationship \citep[e.g.,][]{2014SoPh..289.3949B, 2023JASTP.24205981B}. Identifying optimal functions that define the correlation between FD and IP parameters remains an open research area, as it is crucial for space weather prediction.\\ 
The temporal delay between geomagnetic disturbances and FD onset is a key aspect of their relationship. The storm sudden commencement often coincides with FD onset \citep[e.g.,][]{raghav2021role, 2023JASTP.25206146G}.
Recent studies by \cite{2023JASTP.24205981B} reported a 4-hour delay between solar wind speed and FD, while the time-lag between the IMF southward component Bz and the Dst index was found to be 2 hours \citep[e.g.,][]{Ahmed2024a}.
Previous studies have shown that CR intensity fluctuations linked to strong geomagnetic storms can serve as precursors, allowing forecasts 10--15 hours in advance \citep[e.g.,][]{2005AnGeo..23.2997D}.
\cite{2019SpWea..17..487B} examined the CR-geomagnetic disturbance time lag during the 4--10 September 2017 storm, observing a 3--4 hour delay.
Numerous studies have previously been conducted on FDs \citep[e.g.,][]{2021AdSpR..68.4702B} and geomagnetic storms \citep[e.g.,][]{Ahmed2024a, Ahmed2024} using hourly data.
On the other hand, some studies have focused on short time periods \citep[e.g.,][]{2019SpWea..17..487B}, who analyzed the September 2017 event to investigate time lags between CR intensity and geomagnetic activity. They reported a time lag of three to four hours.
In this study, we address the limitations of previous works by analyzing multiple FD and geomagnetic storm events by using high resolution minute data, allowing for a more detailed investigation of their temporal characteristics and interactions across three solar cycles.
This approach provides more precise measurements of temporal lags, offering new insights into the precursor effects of CR variations and improving the existing storm forecast reliability.\\
The FD amplitude is influenced by cutoff rigidity, which depends on the geomagnetic field's shielding properties \citep[e.g.,][]{2014SoPh..289.3949B}. The cutoff rigidity determines the range of CR energies that can penetrate the Earth's atmosphere, with lower-energy particles being more susceptible to modulation by solar wind and IMFs \citep[e.g.,][]{1987P&SS...35.1101G}.
Previous studies have examined factors affecting cutoff rigidity and energy spectra in neutron monitors, including latitude \citep[e.g.,][]{1962JPSJS..17B.400T, 2005JGRA..110.9S20B}, altitude \citep[e.g.,][]{2021SoPh..296..129M}, and rigidity spectra, including the hardness ratio \citep[e.g.,][]{2011ICRC...10..290W, 2012AdSpR..50..725A}.\\
To understand the temporal relationship between FDs and geomagnetic storms, we first calculate their time delays. We then utilize superposed epoch analysis to investigate the temporal evolution of FDs, with a focus on their main-phase effects in connection to various IP parameters. Furthermore, we use correlation analysis to examine the relationship between FD amplitudes, various IP parameters, and the energy spectrum during strong geomagnetic disturbances, aiming to identify the most effective parameter that describes FDs.
\section{Data and Method} \label{data}
\subsection{Data Identification}\label{ident}
\begin{figure*}[t]
	\centering
	\includegraphics[width=\linewidth, height=14cm]{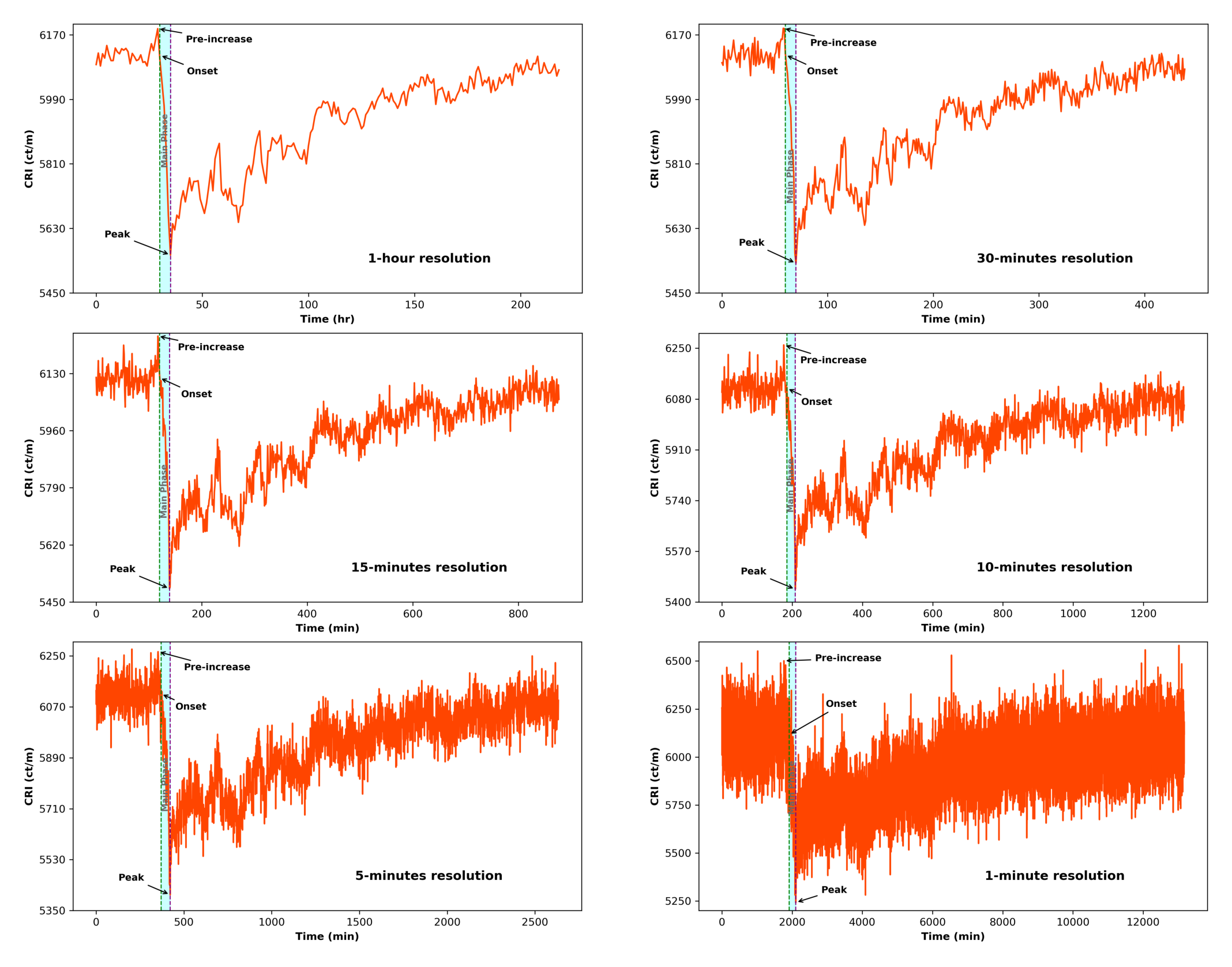}
\caption{Time profiles of the 2005-05-15T01:30 FD event at multiple resolutions: 1 h, 30 min, 15 min, 10 min, 5 min, and 1 min. The cyan-shaded area indicates the main phase of the FD, while the green and purple dashed vertical lines mark the onset and peak times, respectively. Key values such as the pre-increase, onset, and peak are annotated with arrowed labels for clarity. CRI stands for CR intensity and the data obtained from the Oulu neutron monitor, were used to produce this plot. \label{ff1}}
\end{figure*}
We extracted FD events, their associated geomagnetic storms, and various IP parameters for the period 1995 -- 2018, using a minute time-binned data.
We identified FD events based on the following criteria: (a) events with a main-phase amplitude $\geq 1.3\%$; (b) the accompanying geomagnetic storms of SYM-H $<-50$ nT are considered; (c) excluding events with significant zigzag-like dips;
and (d) events with full recovery are considered. Based on these criteria, a total of 57 events were selected.\\
To identify individual events, including driver sources and timings as well as coronal mass ejection propagating through IP space (ICME) transient parameters (such as maximum speed, average speed, transit time, sheath time, and disturbance time), we used the near-Earth ICME catalog\footnote{\url{https://izw1.caltech.edu/ACE/ASC/DATA/level3/icmetable2.htm}} \citep[e.g.,][]{2003JGRA..108.1156C, 2010SoPh..264..189R} and University of Science and Technology of China (USTC) list of ICMEs \citep[e.g.,][]{2016SoPh..291.2419C}\footnote{\url{https://space.ustc.edu.cn/dreams/wind_icmes/index.php}}. Transit speed was determined based on the Sun-Earth distance (1 AU) and transit duration. Flare features (such as halo or non-halo) were obtained from the SOHO-LASCO ICME database\footnote{\url{https://cdaw.gsfc.nasa.gov/CME_list/}}.\\
CR intensity data were obtained from the worldwide neutron monitor database\footnote{\url{https://www.nmdb.eu/nest/}}, 
while IP solar wind parameters and geomagnetic indices were retrieved from the OMNIWeb online tool\footnote{\url{https://omniweb.gsfc.nasa.gov/}}. These data were provided in the geocentric solar magnetospheric (GSM) coordinate system, based on measurements from the ACE and WIND spacecraft. The parameters include geomagnetic index SYM-H (minute resolution), IMF components (total B, fluctuations $\sigma$B, and the south-north component Bz), and IP plasma properties like temperature T, dynamic pressure P, density $\rho$, dawn-dusk electric field Ey, and plasma beta $\beta$.\\
To study FD dynamics, we derived IP parameters ($\sigma$B/B, vB, vB$\rho$, P$\times$(vB), B$\times$v, $\sigma$B$\times$B$\times$v, ($\sigma$B/B)$\times$B$\times$v). The goal of deriving those coupling functions was to identify parameters that best represent the main phase of FDs.\\
Because of the dynamic and variable nature of the solar wind, CR flux in the heliosphere exhibits continuous fluctuations across a wide range of timescales. Consequently, establishing a clear and consistent definition of a FD remains a challenge.
Visual/manual procedure to identify and to calculate the FD magnitude has been commonly employed in earlier studies. However, biases in manually estimated FD magnitudes have been reported. \cite{2020MNRAS.491.3793O} provided with an algorithm (equation 1) to calculate FD magnitude.
\cite{2020ApJ...896..133L} has established an automated method of FD identification. More recently, \cite{2024A&A...683A.168D} suggested a best-fit method for measuring FDs, which performs similarly to the algorithm-based method, but with a slightly smaller value of magnitude. In view of the limitations of individual methods, \cite{2020MNRAS.491.3793O} suggested a combination of machine and manual technique to tackle the problem of FD identification.
In this study we used visual identification (by plotting the cosmic ray time profile) together with a manual calculation of FD magnitude (using  Equation \ref{eq1}). To start with,  chart is generated at (\url{https://cosmicrays.oulu.fi}) for selected periods. Visual identification of these charts helps us to select FD events with mentioned criteria. It provides us with an idea about  the shape of  FDs, their onset, minimum time and recovery profile as well as the magnitude. Then a  manual calculation (using pressure corrected data of neutron monitors)  leads us to determine the actual amplitude of individual FDs (Equation \ref{eq1}). To account for the noise in the data and to define a uniform criteria for the pre-decrease, we have taken the average of the 48 hour count rate prior to onset of the FD. This value is taken as CRF$_{\mathrm{max}}$  in Equation \ref{eq1} for the calculation of FD magnitude.
\subsubsection{Manual identification}\label{man}
We applied a manual FD identification method \citep[e.g.,][]{1997ICRC....2..197P, 2008JGRA..113.1103O, 2020MNRAS.491.3793O, 2024A&A...683A.168D} for visual inspection of CR intensity variations. By observing the CR intensity fluctuations, we can readily identify the general shape of an FD event namely, the initial, main, and recovery phases by eye. However, accurately determining the onset time remains challenging using this method. As shown in Figure \ref{ff1}, the temporal variation of the CR flux reveals the distinct phases of the FD event.
\subsubsection{Automated identification}\label{aut}
Once the FDs were identified manually, we calculated their amplitudes using the method described by \citep[][Equation 1]{2020ApJ...896..133L}, and expressed it as a percentage by multiplying the result by 100.
However, \cite{2025JApA...46....8E} utilized an automated FD location algorithm (Equation 1) applied to daily-averaged CR data.
Next we calculated the FD amplitude as the relative drop in CR count rate, defined by:
\begin{equation}
\mathrm{FD_{amplitude}=(\frac{CRF_{max}-CRF_{min}}{CRF_{max}})\times100}\label{eq1}
\end{equation}
where $\mathrm{CRF_{max}}$ is the onset CR flux (background level) and $\mathrm{CRF_{min}}$ is the CR flux at the event's minimum \citep[e.g.,][]{2011A&A...531A..91D}.
In Figure \ref{ff1}, we present data from a single neutron monitor\footnote{\url{https://cosmicrays.oulu.fi/}}, as our objective is to illustrate the morphology (i.e., distinct phases) throughout the temporal variation of FD events.
\subsection{Data Classification}\label{class}
\begin{figure*}[t]
	\centering
	\includegraphics[width=\linewidth, height=13.5cm]{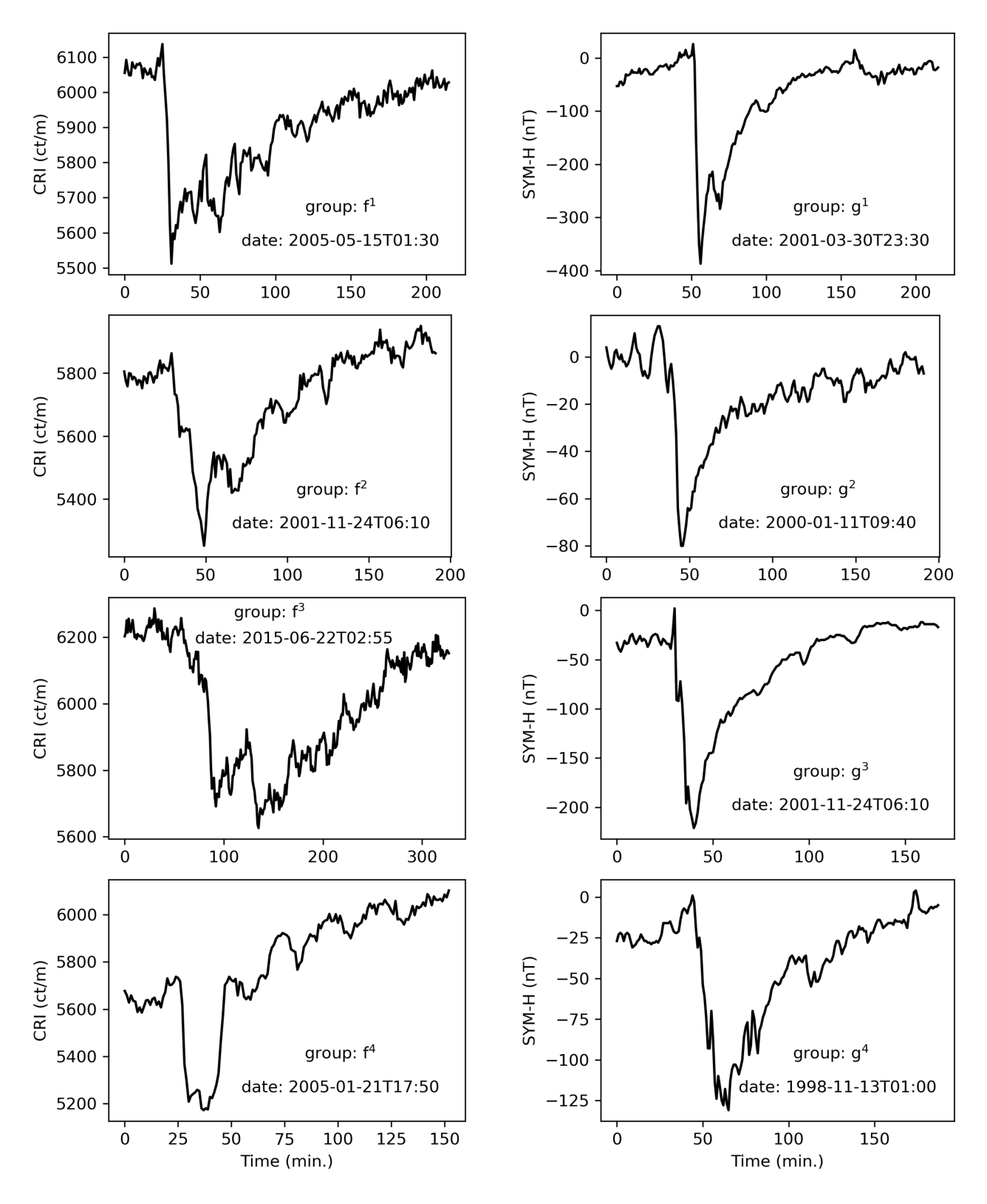}
 \caption{Time variations representing each group of FD events and their associated geomagnetic storms. The first column (top to bottom) displays f$^1$, f$^2$, f$^3$ and f$^4$, while the second column (top to bottom) shows g$^1$, g$^2$, g$^3$ and g$^4$; both are indicated by black lines. See Section \ref{class} for a comprehensive description of the grouping procedure. Data were obtained from the Oulu neutron monitor.\label{f1}}
\end{figure*}

The selected FD and geomagnetic storm events were categorized into groups according to the steps observed during their main phase decrease. This classification aids in understanding the main phase dynamics of each group, particularly in relation to IP solar wind parameters. Additionally, it provides insights into which groups of events are more likely to generate significant disturbances in Earth's magnetosphere.
FD events were classified into five categories based on the number of distinct reduction steps observed in their main phases, from onset to peak values:
\begin{itemize}
\item One-step (f$^1$): Events characterized by a single-step reduction in CR intensity 
\item Two-step (f$^2$): Events exhibiting two distinct reductions 
\item Three-step (f$^3$): Events with three successive reductions in CR intensity 
\item Complex (f$^4$): Events displaying intricate dynamics in their reduction and recovery patterns 
\item CIR-driven (f$^\mathrm{CIR}$): A special category of events triggered by corotating interaction regions.
\end{itemize}
Geomagnetic storms were categorized into four groups based on the number of distinct reduction steps observed during the main phase of the disturbance, as follows:
\begin{itemize}
\item g$^1$: Storms exhibiting a single-step decrease 
\item g$^2$: Storms with two distinct decreases 
\item g$^3$: Storms characterized by three successive decreases 
\item g$^4$: Storms displaying multiple-step decreases 
\end{itemize}
Figure \ref{f1} shows the time variations of events as representative for each group of FD events\footnote{See Section \ref{class} for grouping details and explanations.} and their corresponding geomagnetic storms. In the first column (top to bottom), f$^1$, f$^2$, f$^3$ and f$^4$, whereas in the second column (top to bottom), g$^1$, g$^2$, g$^3$ and g$^4$  both are represented by black lines.
\subsection{Analysis}\label{analys}
We analyzed the classified FDs and geomagnetic storms as well as IP solar wind data
utilizing minute time bins.
For FDs, the pre-increase and pre-decrease phases were isolated \citep{lingri2016forbush}, and for geomagnetic storms, storm sudden commencements were identified prior to the onset carefully. To mitigate noise present in high-resolution data, we analyzed the FD event using smoothing techniques like average superposed analysis with progressively finer time bins 1 hour, 30 minutes, 15 minutes, 10 minutes, 5 minutes, and 1 minute as illustrated in Figure \ref{ff1}. In our analysis, statistical noise is likely the predominant factor contributing to the observed fluctuations in high-temporal-resolution data. This approach is frequently employed in space weather studies to optimize both signal resolution and time-domain precision \citep[e.g.,][]{2019SpWea..17..487B}.
The onset of FD in the employed superposed epoch analysis (as shown in Figures \ref{fig:3}, \ref{fig:7}, \ref{fig:4}, \ref{fig:5}, \ref{fig:6}) were determined through equation \ref{eq2}, as we took two days (2880 minutes) initial phase for 1 minute time bins.
\\
The two-days initial phase and onset were determined by averaging the count rate values over the given time intervals: 2880 minutes.
\begin{equation}
 \mathrm{CRI_{mean}=\frac{Two~day ~CR ~data ~during~ inital~ phase}{2880~ minutes}}\label{eq2}
\end{equation}
This mean value ($\mathrm{CRI_{mean}}$) was used to define the FD onset. Unlike CR data, geomagnetic storm data is less noisy, which simplified the SSC peak value and storm onset identification.
In addition, to calculate time lags between FDs and geomagnetic storms \citep{2023JASTP.24205981B}, we used those methods.
High-resolution minute data were used to pinpoint these timings more precisely.
A superposed epoch analysis was employed for each group to observe the average time variation between FDs and IP parameters \citep[e.g.,][]{2021Ap.....64..426F, 2021AdSpR..68.4702B}. This method is useful to understand the average dynamics of time variation between FD and IP transient parameters. The initial phase was defined as the two days preceding the onset in the superposed plots, during which the pre-increase and pre-decrease values were carefully examined. Following the initial phase, the main phase of FDs begins and is represented by a gray area in the superposed plots. \\
In addition to superposed analysis, correlation studies were conducted to determine relationships between FD amplitude and peak values of (geomagnetic indices, IP parameters), ICME transient properties, cutoff rigidity, and energy spectra.
The relationship between FD amplitude and peak IP parameters using single, double, and triple parameter combinations were analyzed attempting to identify best coupling function which best represent the FD dynamics.
The correlation between FD amplitude and the properties of SYM-H (minimum and amplitude) was also examined for CME-triggered events.\\
The relationship between FD amplitude and ICME transients was analyzed based on their manifestations in IP space, such as ejection extent or size, speed, and timing of the ICME as discussed in Section \ref{transient}.
Finally, we examined the energy dependency of FDs using correlation analysis utilizing geomagnetic cutoff rigidity and median energy data from several neutron monitor stations as explained in Section \ref{cutoff}.
\section{Results and Discussion} \label{res}
\subsection{Time-lag between FD and geomagnetic storm}\label{sub:lag}
\begin{table*}
	\centering
	\caption{List of the onset of strong magnetic storms, accompanying FDs, and time-lag between events. The time lag/lead (+/-) is defined as follows: \textquotedblleft +\textquotedblright when FD leads SYM-H, while \textquotedblleft-\textquotedblright when FD lags SYM-H.}\label{t2}
	\begin{tabular}{@{}lccccccc@{}}
		\toprule
		\multicolumn{2}{@{}c@{}}{SYM-H}&\multicolumn{3}{@{}c@{}}{FD}&\multicolumn{2}{@{}c@{}}{Time difference}&Time \\
		Start date & gp & Start Date&End date &gp&hour&minute&lag/lead \\
		\midrule
		1995-04-07T01:10 & g$^3$ & 1995-04-07T01:40 & 1995-04-14T15:00 & f$^{\mathrm{CIR}}$&00&30&-\\
		1995-10-18T12:00 & g$^3$ & 1995-10-18T07:10 & 1995-10-26T07:00 & f$^4$&04&50&+\\
		1997-01-10T02:28 & g$^1$ & 1997-01-10T02:35 & 1997-01-15T10:00 & f$^2$&00&07&-\\
		1997-04-21T09:30 & g$^2$ & 1997-04-21T10:20 & 1997-04-23T06:00 & f$^3$&00&50&-\\
		1997-05-01T15:05 & g$^3$ & 1997-05-01T13:22 & 1997-05-02T09:00 & f$^{\mathrm{CIR}}$&01&43&+\\
		1997-05-15T06:15 & g$^3$ & 1997-05-15T06:40 & 1997-05-15T23:00 & f$^1$&00&25&-\\
		1997-05-26T11:22 & g$^3$ & 1997-05-26T10:35 & 1997-05-28T14:00 & f$^3$&00&47&+\\
		1997-06-07T01:00 & g$^4$ & 1997-06-07T02:20 & 1997-06-18T09:00 & f$^1$&01&20&-\\
		1997-09-03T16:26 & g$^4$ & 1997-09-03T09:40 & 1997-09-05T09:00 & f$^3$&06&46&+\\
		1997-10-10T18:20 & g$^2$ & 1997-10-10T14:15 & 1997-10-13T08:00 & f$^2$&04&05&+\\
		1997-11-06T22:52 & g$^1$ & 1997-11-06T16:50 & 1997-11-10T10:00 & f$^1$&06&02&+\\
		1998-01-06T15:00 & g$^2$ & 1998-01-06T11:13 & 1998-01-16T06:00 & f$^3$&03&47&+\\
		1998-03-10T10:37 & g$^2$ & 1998-03-10T11:37 & 1998-03-16T09:00 & f$^{\mathrm{CIR}}$&01&00&-\\
		1998-06-25T22:50 & g$^2$ & 1998-06-25T18:40 & 1998-06-27T07:00 & f$^4$&04&10&+\\
		1998-08-06T01:00 & g$^3$ & 1998-08-05T19:00 & 1998-08-07T05:00 & f$^3$&04&00&+\\
		1998-08-26T12:50 & g$^4$ & 1998-08-26T10:50 & 1998-08-30T09:00 & f$^3$&02&00&+\\
		1998-09-25T00:30 & g$^3$ & 1998-09-24T18:55 & 1998-10-01T07:00 & f$^3$&06&35&+\\
		1998-11-13T00:45 & g$^4$ & 1998-11-13T01:00 & 1998-11-17T13:00 & f$^3$&00&15&-\\
		1999-01-13T02:20 & g$^4$ & 1999-01-12T17:40 & 1999-01-22T05:00 & f$^4$&05&40&+\\
		1999-09-22T20:10 & g$^2$ & 1999-09-22T16:10 & 1999-09-23T06:00 & f$^4$&04&00&+\\
		2000-01-11T08:30 & g$^2$ & 2000-01-11T09:40 & 2000-01-17T15:00 & f$^{\mathrm{CIR}}$&01&10&-\\
		2000-04-06T16:40 & g$^1$ & 2000-04-06T12:20 & 2000-04-09T12:00 & f$^3$&04&20&+\\
		2000-07-15T19:10 & g$^3$ & 2000-07-15T12:15 & 2000-07-19T12:00 & f$^2$&06&55&+\\
		2000-08-12T03:30 & g$^1$ & 2000-08-12T03:35 & 2000-08-13T06:00 & f$^4$&00&05&-\\
		2000-09-17T19:53 & g$^1$ & 2000-09-17T12:40 & 2000-09-24T09:00 & f$^2$&07&13&+\\
		2001-03-31T04:10 & g$^1$ & 2001-03-30T23:30 & 2001-04-04T09:00 & f$^3$&04&40&+\\
		2001-04-11T15:45 & g$^1$ & 2001-04-11T16:50 & 2001-04-19T12:00 & f$^3$&01&05&-\\
		2001-04-18T00:48 & g$^1$ & 2001-04-18T01:05 & 2001-04-20T13:00 & f$^4$&00&17&-\\
		2001-04-22T00:00 & g$^3$ & 2001-04-21T20:45 & 2001-04-23T15:00 & f$^4$&03&15&+\\
		2001-08-17T16:50 & g$^1$ & 2001-08-17T16:20 & 2001-08-20T21:00 & f$^1$&00&30&+\\
		2001-10-03T06:50 & g$^1$ & 2001-10-03T07:00 & 2001-10-04T07:00 & f$^4$&00&10&-\\
		2001-11-05T18:55 & g$^3$ & 2001-11-05T19:10 & 2001-11-12T12:00 & f$^2$&00&15&-\\
		2001-11-24T06:45 & g$^3$ & 2001-11-24T06:10 & 2001-11-28T13:00 & f$^2$&00&35&+\\
		2002-09-07T13:20 & g$^2$ & 2002-09-07T13:50 & 2002-09-13T15:00 & f$^3$&00&30&-\\
		2003-11-20T03:30 & g$^4$ & 2003-11-20T04:10 & 2003-11-21T15:00 & f$^1$&00&40&-\\
		2004-04-03T14:40 & g$^2$ & 2004-04-03T10:45 & 2004-04-08T03:00 & f$^3$&03&55&+\\
		2004-08-30T01:05 & g$^4$ & 2004-08-30T03:20 & 2004-09-05T13:00 & f$^4$&02&15&-\\
		2005-01-21T19:45 & g$^4$ & 2005-01-21T17:50 & 2005-01-22T16:00 & f$^4$&01&55&+\\
		2005-05-15T06:15 & g$^1$ & 2005-05-15T01:30 & 2005-05-22T07:00 & f$^1$&04&45&+\\
		2006-12-14T21:30 & g$^4$ & 2006-12-14T14:50 & 2006-12-18T13:00 & f$^1$&06&40&+\\
		2009-07-21T22:17 & g$^2$ & 2009-07-21T23:19 & 2009-07-23T16:00 & f$^4$&01&02&-\\
		2011-08-05T20:20 & g$^2$ & 2011-08-05T17:05 & 2011-08-12T14:00 & f$^2$&03&15&+\\
		2011-10-24T22:10 & g$^2$ & 2011-10-24T15:50 & 2011-10-30T08:00 & f$^1$&06&20&+\\
		2012-04-23T17:40 & g$^4$ & 2012-04-23T17:35 & 2012-04-29T04:00 & f$^4$&00&05&-\\
		2012-07-15T01:15 & g$^4$ & 2012-07-14T17:50 & 2012-07-18T15:00 & f$^2$&07&25&+\\
		2012-11-12T19:20 & g$^4$ & 2012-11-12T20:00 & 2012-11-17T15:00 & f$^4$&00&40&-\\
		2013-03-17T07:20 & g$^4$ & 2013-03-17T03:20 & 2013-03-25T09:00 & f$^4$&04&00&+\\
		\botrule
	\end{tabular}
\end{table*}
\begin{table*}
	\centering
	\begin{tabular}{@{}lccccccc@{}}
		\toprule
	\multicolumn{2}{@{}c@{}}{SYM-H}&\multicolumn{3}{@{}c@{}}{FD}&\multicolumn{2}{@{}c@{}}{Time difference}&Time \\
	Start date & gp & Start Date&End date &gp&hour&minute&lag/lead \\
		\midrule
		2013-06-01T00:45 & g$^2$ & 2013-05-31T20:50 & 2013-06-05T10:00 & f$^{\mathrm{CIR}}$&03&55&+\\
		2013-06-06T15:30 & g$^4$ & 2013-06-06T11:40 & 2013-06-12T06:00 & f$^3$&03&50&+\\
		2014-02-27T18:15 & g$^3$ & 2014-02-27T19:00 & 2014-03-07T11:00 & f$^{\mathrm{CIR}}$&00&45&-\\
		2015-01-07T08:10 & g$^1$ & 2015-01-07T08:20 & 2015-01-08T11:00 & f$^4$&00&10&-\\
		2015-03-17T06:40 & g$^3$ & 2015-03-16T23:40 & 2015-03-25T08:00 & f$^3$&07&00&+\\
		2015-06-22T05:45 & g$^4$ & 2015-06-22T02:55 & 2015-07-02T10:00 & f$^3$&02&50&+\\
		2015-12-31T11:20 & g$^4$ & 2015-12-31T09:55 & 2016-01-06T08:00 & f$^2$&01&25&+\\
		2016-10-13T02:15 & g$^1$ & 2016-10-12T23:05 & 2016-10-23T01:00 & f$^4$&03&10&+\\
		2017-05-27T22:45 & g$^1$ & 2017-05-27T18:35 & 2017-06-08T09:00 & f$^3$&04&10&+\\
		2018-08-25T13:30 & g$^4$ & 2018-08-25T07:50 & 2018-08-26T03:00 & f$^4$&07&40&+\\
		\botrule
		\multicolumn{7}{l}{\footnotesize gp -- stands for group}
	\end{tabular}
\end{table*}

\begin{figure*}[t]
	\centering
	\includegraphics[width=0.88\linewidth, height=10.5cm]{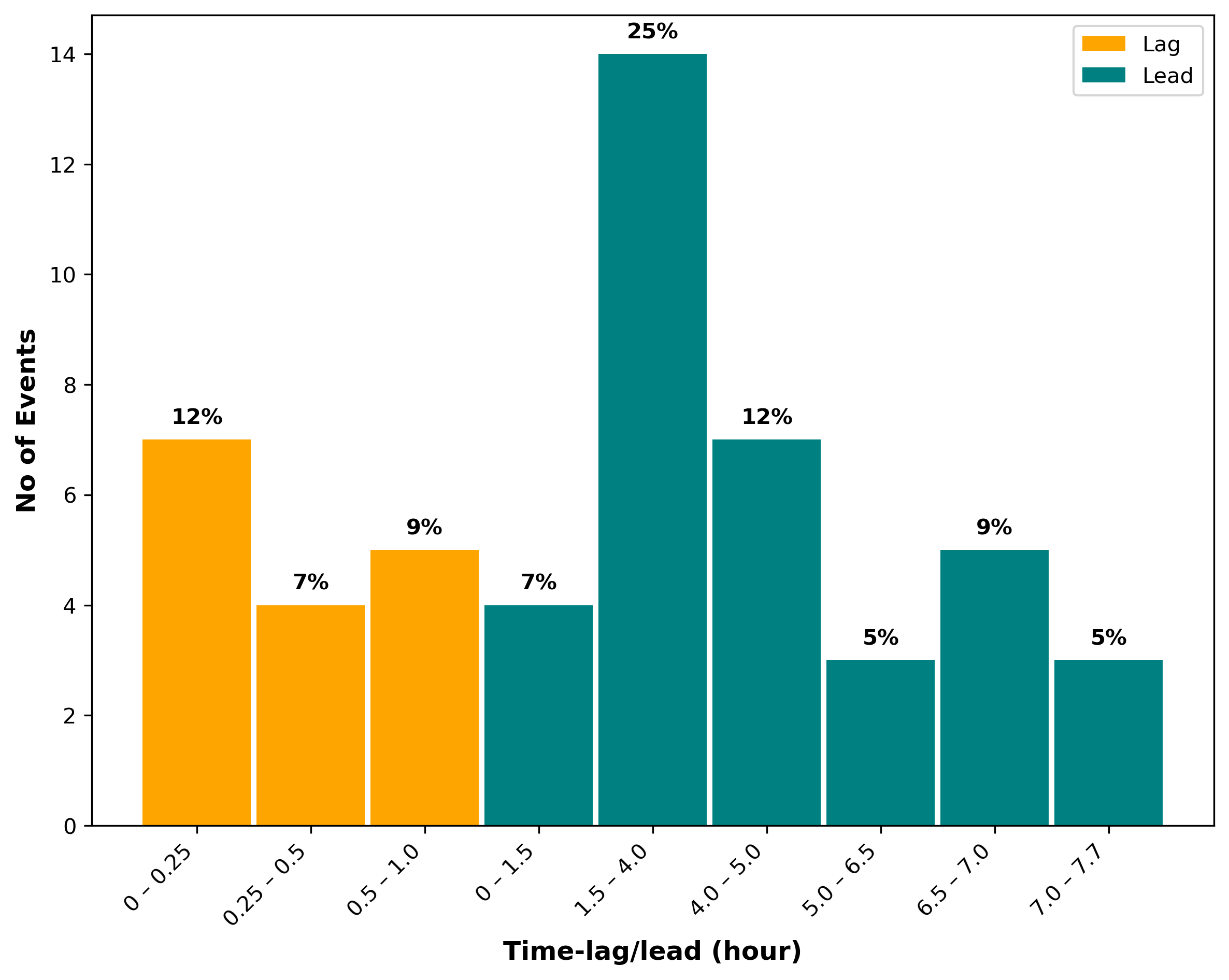}
\caption{Histogram of time-lag between FDs and magnetic storms based on main phase onset time. The orange colored blocks show time lag events and the teal blocks show time lead events within each time bin.
		\label{fig:01}}
\end{figure*}
\begin{table*}
	\centering
\caption{Distribution of events by time-lag and time-lead intervals.}\label{t22}
\begin{tabular}{lcc|ccc}
		\toprule
		\multicolumn{3}{c|}{Time lag} & \multicolumn{3}{c}{Time lead} \\
		Interval & No of & Percentage & Interval & No of & Percentage \\
		hours & events & (\%) & hours & events & (\%) \\
		\midrule
		0 -- 0.25 & 7 & 12 & 0 -- 1.5 & 4 & 7 \\
		0.25 -- 0.5 & 4 & 7 & 1.5 -- 4.0 & 14 & 25 \\
		0.5 -- 1.0 & 5 & 9 & 4.0 -- 5.0 & 7 & 12 \\
		1.0 -- 1.5 & 5 & 9 & 5.0 -- 6.5 & 3 & 5 \\
		 &  & & 6.5 -- 7.0 & 5 & 9 \\
		 & & & 7.0 -- 7.7 & 3 & 5 \\
		 \hline
		Total&21 &37 & Total & 36 & 63 \\
		\bottomrule
	\end{tabular}
\end{table*}
We have divided the time-lag and time-lead intervals into separate time blocks. The time-lag intervals are (0 -- 0.25, 0.25 -- 0.5, 0.5 -- 1.0, 1.0 -- 1.5) hours, accounting for 12\%, 7\%, 9\%, and 9\% of the events, respectively, with a total of 21 events. The time-lead intervals are (0 -- 1.5, 1.5 -- 4.0, 4.0 -- 5.0, 5.0 -- 6.5, 6.5 -- 7.0, 7.0 -- 7.7) hours, corresponding to 7\%, 25\%, 12\%, 5\%, 9\%, and 5\% of the events, respectively, totaling 36 events.
Figure \ref{fig:01} and Table \ref{t22} shows the time-lag/lead (-/+) between the onset of FDs and geomagnetic disturbances. We have assigned time-lag (-) is when geomagnetic storm starts earlier than FD and time-lead (+) is when FD begins earlier than geomagnetic disturbance onset time.
The reader should bear in mind that minute-resolution data were used to compute the time lag, but the time axis is shown in hours for clarity and space efficiency. \\
Studying the time-lag/lead based on the onset time followed by sudden decrease of FDs and magnetic storms is crucial for space weather prediction and forecasting. Our findings show that the time lag/lead between geomagnetic disturbances and FDs start time takes from 5 minutes to nearly eight hours as shown from Table \ref{t2}.
Table \ref{t2} and Figure \ref{fig:01} show that 25\% of events have a 1.5 -- 4.0 hours (90 -- 240 minutes) lead, corresponding to 1.4 -- 9.0\% FD amplitude. This is consistent with a previous study by \cite{2019SpWea..17..487B}, who found that the geomagnetic storm and FD have time lags of 1 -- 4 hours in their multi-time bin observations of ICMEs. \cite{2011A&A...531A..91D} studied the onset time lag between FD and the IMF B and reported more than 3 hours. \cite{2023JASTP.24205981B} reported that approximately 4 hours delay between solar wind speed and FD from their wavelet analysis. According to \cite{2023JASTP.25206146G}, the only temporal lag between the decrease in CR intensity and the onset of geomagnetic disturbance is the storm sudden commencement\footnote{Strong IP shocks compress the magnetosphere, causing most moderate and powerful storms to begin quickly \citep[e.g.,][]{1992GeoRL..19.1227R}. Fast CMEs are often the predominant cause of storm sudden commencements \citep[e.g.,][]{1992GeoRL..19.1227R}.
The magnetopause compresses as the solar wind approaches the bow of the magnetosphere, resulting in a rapid increase in the magnetic field on the Earth's dayside, which lasts only a few minutes \citep[e.g.,][]{OYEDOKUN2018421}.} prior to the main phase magnetic storm.\\
Despite the fact that the mechanisms by which the two events (FDs and geomagnetic disturbances) develop are distinct, $~9\%$ which corresponds to five events have lied down in the same group (e.g., g$^1\sim$ f$^1$ and g$^2\sim$ f$^2$). We have excluded the f$^{\mathrm{CIR}}$ group from this calculation since we did not consider its group in the geomagnetic disturbances.
Similarly, g$^3\sim$ f$^3$ and g$^4\sim$ f$^4$ were excluded from consideration because their grouping criteria for FD events and geomagnetic storms differ from those outlined in Section \ref{class}.
\subsection{Superposed epoch analysis and time variations}\label{sub:sea}
Superposed epoch analysis has been used to study the average behavior of geomagnetic disturbances \citep[e.g.,][]{1994AnGeo..12..612T, 2010AGUFMSM23B..08H, 2011JGRA..116.9211H, 2016Ap&SS.361..253B, 2018SoPh..293..126P, Ahmed2024a}. Superposed epoch analysis has also been applied to analyze FDs \citep[e.g.,][]{2006JASTP..68..803S, 2021AdSpR..68.4702B, 2024SoPh..299...16T}. These analyses explore the connection between FDs and IP solar wind plasma and field parameters, which are primarily influenced by solar drivers such as CMEs.
We used the superposed epoch analysis to correlate average FD values with IP solar wind parameters as well as derived functions for minute time bins for each group (f$^1$, f$^2$, f$^3$, f$^4$ and f$^{\mathrm{CIR}}$) (Figures \ref{fig:3} and \ref{fig:7}, as well as \ref{fig:4}, \ref{fig:5} \& \ref{fig:6} in Appendix \ref{A2}).
Each figure comprises of 13 panels from top to bottom: CR intensity, $\sigma$B, B, Bz, $\sigma$B/B, T, $\rho$, v, P, Ey, $\beta$, vB and vB$\rho$, each with various colors for minute resolution. The derived functions vB and vB$\rho$  have a multiplication order of 10$^{-3}$ or 10$^{-3}\times$vB and 10$^{-3}\times$vB$\rho$. The zero hour represents the onset time, while the light-gray region highlights the period from the onset to the peak time (main phase) of the FD. These superposed figures show that the main phase duration of FD is comparatively proportional to that of magnetic disturbances as compared with the \citep{Ahmed2024a}.\\
Figure \ref{fig:3} shows the averaged superposed epoch plot of FD for a one-step decrease (f$^1$). At the time of the FD onset, the perturbed IMF sigma ($\sigma$B), plasma density ($\rho$), dynamic pressure (P), plasma temperature (T), and the derived function ($10^{-3}\times$vB$\rho$) reach their peaks. This provides clear evidence that a turbulent, high-magnetic-field structure has already formed and is passing through before the FD depression begins. Such behavior is consistent with the findings of \citep[e.g.,][]{1988JGR....93.2511Z}.
In contrast to \cite{2021AdSpR..68.4702B}, who discovered that the high-field turbulent structure peaks coincide with the FD peak, our data show that the high-field turbulent region reaches its peak a few hours earlier.
The main phase takes 7 hours to reach its peak value and its FD amplitude is 4.3\%.
Figure \ref{fig:3} shows that the shock region develops a few hours before the onset of FD depression, as evidenced by the sharp enhancement of plasma (velocity, pressure, density) and IMF (magnitude B \& Sigma $\sigma$B) during sheath time. The CR intensity depression takes place in the sheath's disturbed field region, as indicated by a high $\sigma$B.
The IP solar wind velocity (v), total IMF (B), and IP electric field (vB) begin before FD onset and peak during FD recovery. Temperature fully recovered during the peak of FD. Other IP parameters and derived products peak during the FD main phase period.
Using averaged minute resolution to determine specific commencement time can be clearly visible (e.g., see Figure \ref{fig:3}).\\
Figure \ref{fig:7} is the same as Figure \ref{fig:3} and depicts CIR-driven FDs. During FD onset, the IP parameter $\sigma$B reaches its peak. This figure's morphology is identical to Figure \ref{fig:3}, except the main phase lasts 41 hours and has a 1.8\% FD amplitude. IP parameters ($\sigma$B, Bz, $\rho$, Ey) recovered earlier than the peak time of FD. Except for velocity, which extends almost nine days to return to normal level, the other IP parameters return to a quiet level a few hours after the FD peak. Except for $\sigma$B, $\rho$, and P, all parameters show gradual enhancement, indicating no shock/sheath generation prior to FD for CIR events.\\
Figure \ref{fig:4} is the same as Figure \ref{fig:3}, which shows a two-step decrease in FD, where the first step takes $\sim70\%$ of the main phase, with $\sim$96\% of the amplitude decreases during the first-step and the second is steady and takes short time. All IP and plasma parameters are in recovering during the first step decline. The turbulent high field structure peaks and recovered during the first step decrease, while in the second step decrease non-turbulent magnetic cloud or ICME (ejecta) are passing. At the commencement of FD, IP parameters, $\sigma$B, $\rho$, P, and T are at their highest levels. This figure has a similar morphology to Figure \ref{fig:3}, but the main phase lasts longer (22 hours) and the FD amplitude is 5.4\%. The IP plasma temperature, the north-south component of the IMF (Bz), and the dawn-dusk electric field (Ey) all recovered completely during the main phase of the FD. The solar wind velocity takes more than 5 days to return to its onset level, and it recovers the slowest of all parameters. After the FD peak, the remaining IP parameters return to normal levels within a few hours.\\
Figure \ref{fig:5} is similar to Figure \ref{fig:3} but depicts a three-step decrease in FD. The first, second, and third steps occur over equal durations. However, the first step is very fast, the second is fast, and the third step is slow. Approximately 55\% of the amplitude decrease occurs during the first step.
Except for the IMF total B, plasma speed (v), and the function vB, all IP parameters begin to recover during the first-step decrease. This suggests that a turbulent high-field structure is passing during this period. In contrast, the second and third-step decreases indicate the passage of a high-speed, quiet-field structure associated with a magnetic cloud or ejecta, characterized by a relatively lower magnitude of IMF B.
At the commencement of FD, IP parameters Bz, P, Ey, T, and the associated function 10$^{-3}\times$vB$\rho$ reach their peak values. This figure has comparable morphology to Figure \ref{fig:3}, but with a longer main phase (32 hours) and a 3.4\% FD amplitude. The recovery of plasma density and IMF perturbation occurs before the initiation of FD, indicating that the high density turbulent structure peaks and begins to recover during sheath period before FD.
During the main phase of FD, the IP parameters $\sigma$B, Bz, $\rho$, P, and Ey recovered completely. Solar wind velocity recovers slowly compared to other characteristics, taking around 10 days to return to its onset level. The remaining IP parameters return to a calm level some hours after the FD peak.\\
Figure \ref{fig:6} is identical to Figure \ref{fig:3}, showing the complex shape of FD. Its average main phase shows two-step like decrease with the first fast and the later slow which is more likely similar to Figure \ref{fig:4}.
During the commencement of FD, the IP parameters $\sigma$B, P, T and 10$^{-3}\times$vB$\rho$ reach their peak. This figure has similar morphology to Figure \ref{fig:6}, however the main phase lasts 14 hours and the FD amplitude is 2.4\%. However, $\sim70\%$ of the amplitude is reduced in the first fast step decrease. IP plasma T, IMF Bz, and dawn-dusk electric field Ey entirely recovered before the peak of FD. The IP parameters recover to normal levels within a few hours following the FD peak, with the exception of velocity, which can take up to five days.\\
In general, Figures \ref{fig:3}, \ref{fig:4}, \ref{fig:5} and \ref{fig:6} are CME-triggered FDs that exhibit almost similar dynamics, even though the main phase duration varies among groups. A highly turbulent magnetic field shock/sheath structure formed before the FD onset and passed through during its commencement, as evidenced by a sudden increase in IP plasma density, IMF (B, $\sigma$B) and speed of the solar wind plasma. CIR-driven FD events exhibit delayed amplification of the IP field and plasma parameters, resulting in severely perturbed morphology. This result is evidenced by that the fast high magnetic field structure was not initiated in shock region, rather this structure arrived roughly one day after the FD onset beginning.
Peak values of FD, along with IP field and plasma parameters, as well as derived functions from each group, are listed in Table \ref{tab3}. These values are taken from the superposed results of (Figures \ref{fig:3} -- \ref{fig:6}) for minute time bins.
\begin{figure}[H]
	\centering
	\includegraphics[width=0.99\linewidth, height=9.4cm]{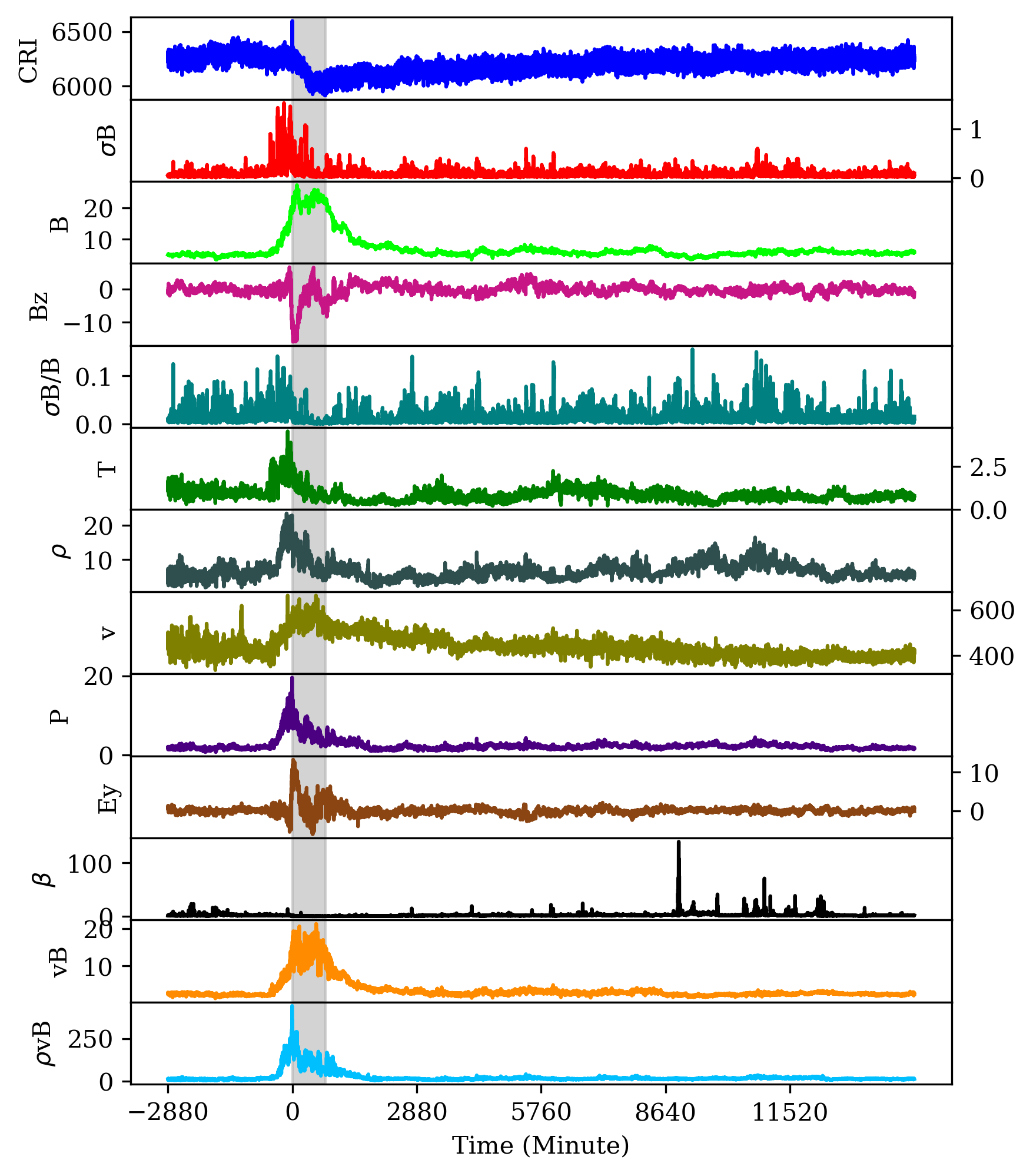}
\caption{Superposed epoch of group f$^1$ of the FD's minute resolution together with the IP parameters, [CR intensity (cts/min), $\sigma\mathrm{B}$ (nT), B (nT), Bz (nT), $\sigma\mathrm{B}$/B, T (K), $\rho$ ($\mathrm{cm}^{-3}$), $\mathrm{v}$ ($\mathrm{kms}^{-1}$), P (nPa), $\mathrm{Ey}$ $(\mathrm{mVm}^{-1})$, $\beta$, $\mathrm{vB}$ $(\mu\mathrm{Tms}^{-1})$, $\mathrm{vB}\rho$ $(\mathrm{pTms}^{-1})$], where pico is ($\mathrm{p}=10^{-12}$). The last two panels of derived parameters vB and vB$\rho$ are in the order fraction of thousands as $10^{-3}\times$vB and $10^{-3}\times\rho$vB. We have multiply the temperature value by a fraction of $10^{-5}$ only for the visualization purpose. The zero time represents the onset time, while the light-gray region highlights the period from the onset to the peak time (main phase) of the FD. The minute resolution CR data was retrieved from Oulu neutron monitor.} \label{fig:3}
\end{figure}
\begin{figure}[H]
	\centering
	\includegraphics[width=0.99\linewidth, height=9.4cm]{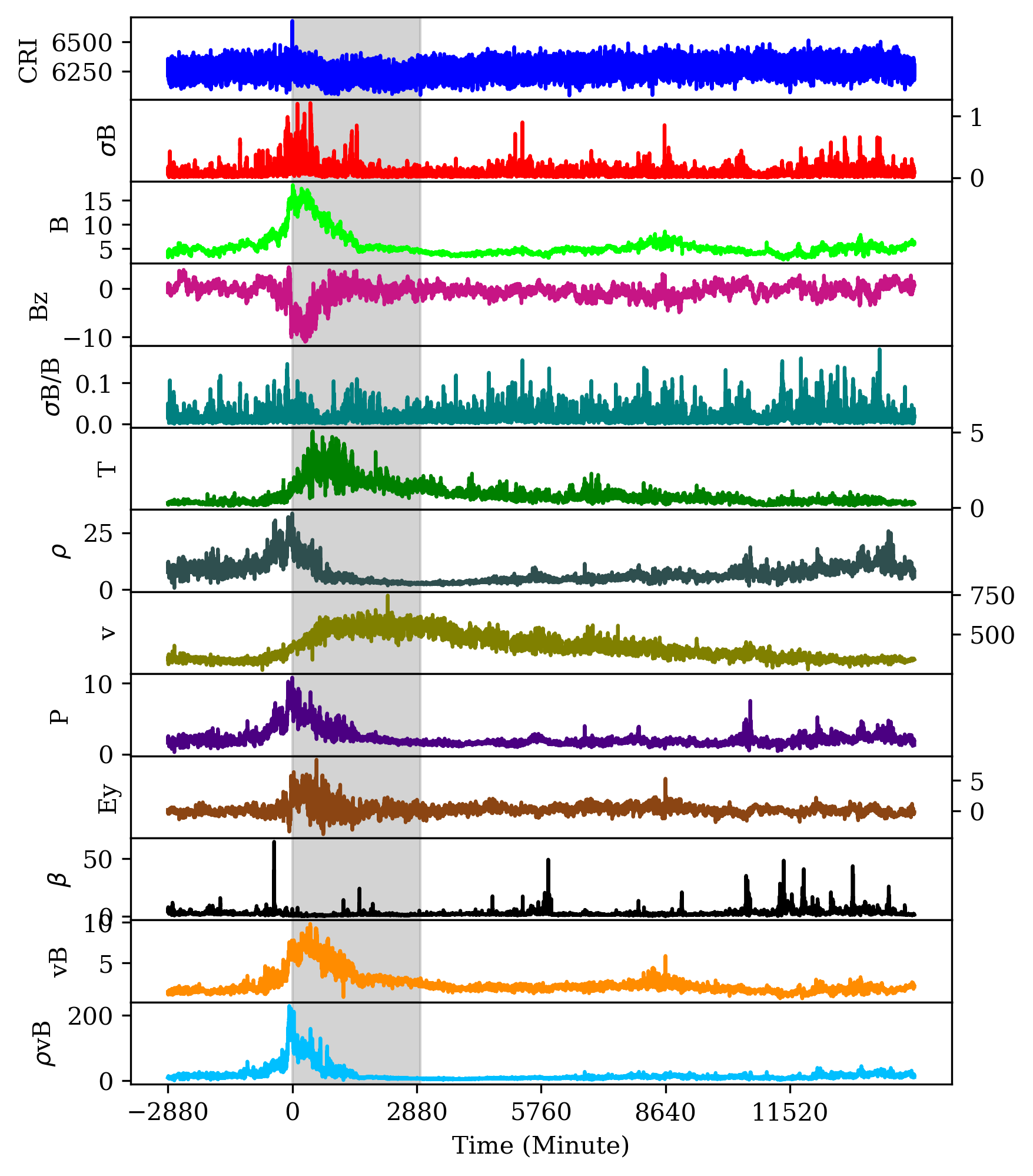}
	\caption{Similar to Figure \ref{fig:3}, except for group f$^{\mathrm{CIR}}$.}
	\label{fig:7}
\end{figure}
\begin{table}
    \caption{Log of FD amplitude in (\%) and peak values of IP parameters as recorded from Figures (\ref{fig:3}, \ref{fig:7}, \ref{fig:4}, \ref{fig:5}, and \ref{fig:6})}\label{tab3}
    \centering
    \begin{tabular}{@{}lccccc@{}}
        \toprule
        Param. & f$^1$ & f$^2$ & f$^3$ & f$^4$ & f$^{\mathrm{CIR}}$ \\
        \midrule
        CR (\%) & 10.4 & 11.9 & 9.0 & 7.8 & 9.4 \\
        B & 27.21 & 24.86 & 19.14 & 17.72 & 18.17 \\
        Bz & -15.93 & -14.27 & -13.74 & -11.30 & -11.01 \\
        $\sigma$B & 1.53 & 1.93 & 1.74 & 0.90 & 1.22 \\
        $\sigma$B/B & 0.14 & 0.17 & 0.11 & 0.10 & 0.16 \\
        T/100000 & 4.54 & 6.69 & 4.56 & 2.27 & 5.04 \\
        $\rho$ & 23.57 & 29.71 & 24.33 & 22.39 & 33.49 \\
        v & 663.35 & 715.37 & 568.33 & 535.59 & 745.8 \\
        P & 19.56 & 25.59 & 14.68 & 13.37 & 10.8 \\
        Ey & 13.18 & 9.95 & 8.95 & 5.81 & 8.33 \\
        $\beta$ & 0.07 & 0.06 & -1.54 & 0.25 & 0.20 \\
        vB/1000 & 21.33 & 19.14 & 11.73 & 9.74 & 9.78 \\
        vB$\rho$/1000 & 441.90 & 517.37 & 315.44 & 205.08 & 228.75 \\
        \botrule
    \end{tabular}
\end{table}

\subsection{Correlation between FD and IP parameters}\label{corr}
\begin{figure*}[ht]
	\centering
	\includegraphics[width=0.76\linewidth]{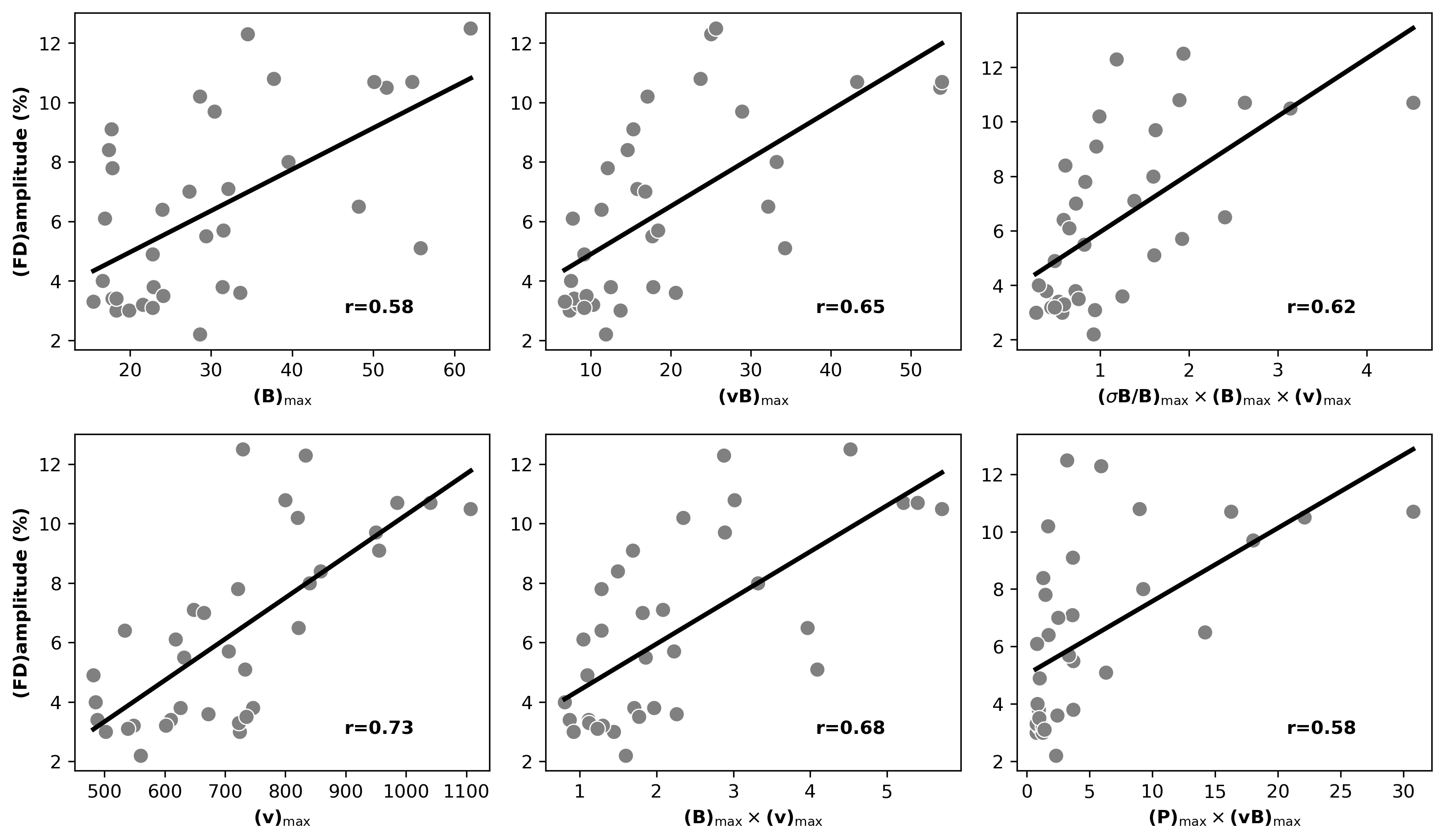}
\caption{The scatter plot for the relationship between the amplitude of FD and the peak values of IP parameters. Two best parameters from each category: one parameter (first column), two parameters (second column), and three parameters (third column) were displayed. The data points (filled circles) are represented by black colors, and the relationship is best described by a linear fit, as shown by the black lines. This analysis utilized CR data taken from Oulu neutron monitor.\label{fig:8}}
\end{figure*}

\begin{figure*}[ht]
	\centering
\includegraphics[width=0.98\textwidth, height=9.4cm]{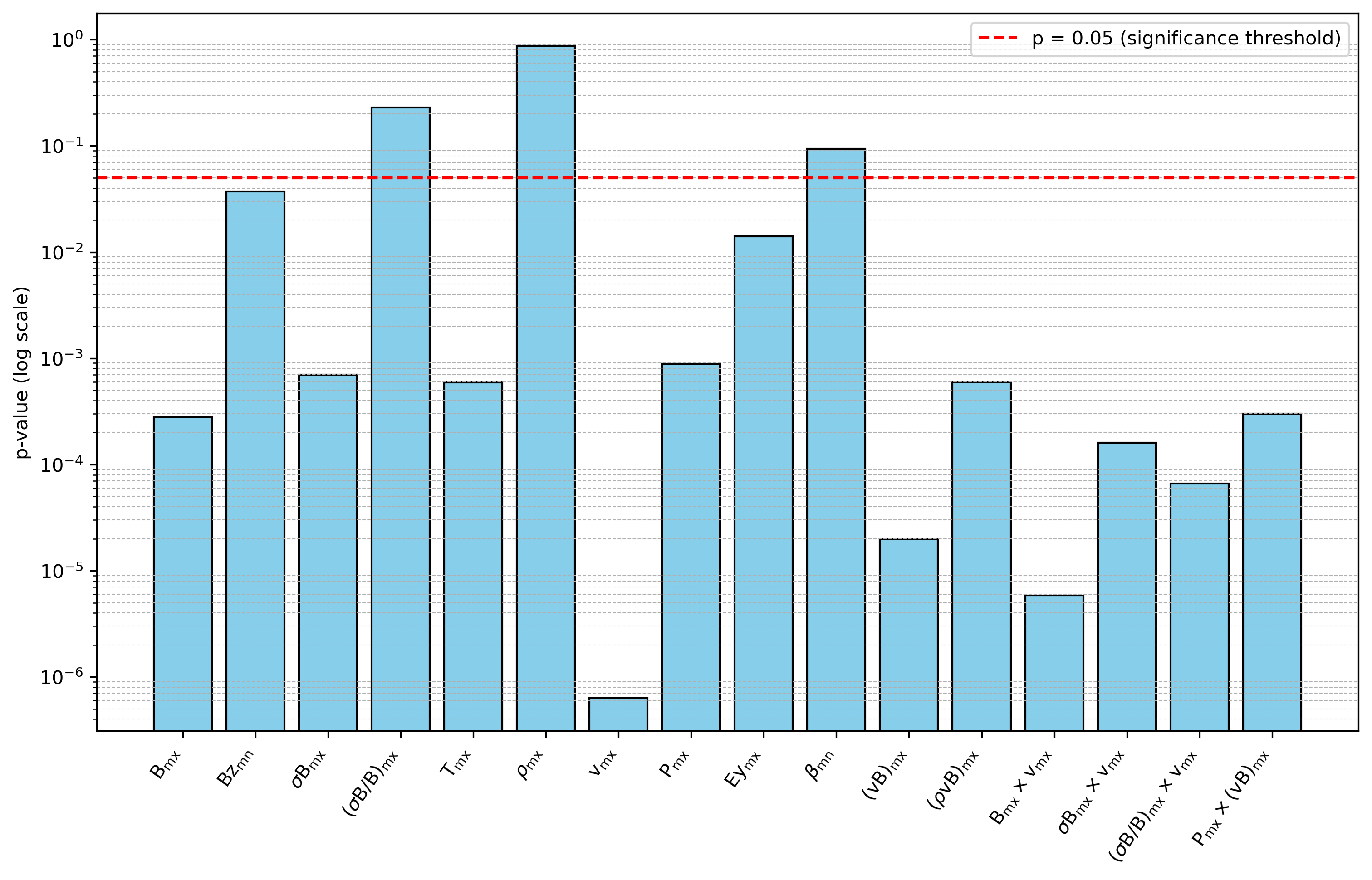}
\caption{Bar chart showing the p-values corresponding to the Pearson correlation between FD magnitude and various IP parameters. Each bar (colored cyan) represents the statistical significance of the correlation for a given parameter. The horizontal dashed red line  marks the threshold for statistical significance (p = 0.05). Parameters with p-values below this line indicate statistically significant correlations.\label{fig:81}}
\end{figure*}

\begin{figure*}[t]
	\centering
	\includegraphics[width=0.37\linewidth, height=4.3cm]{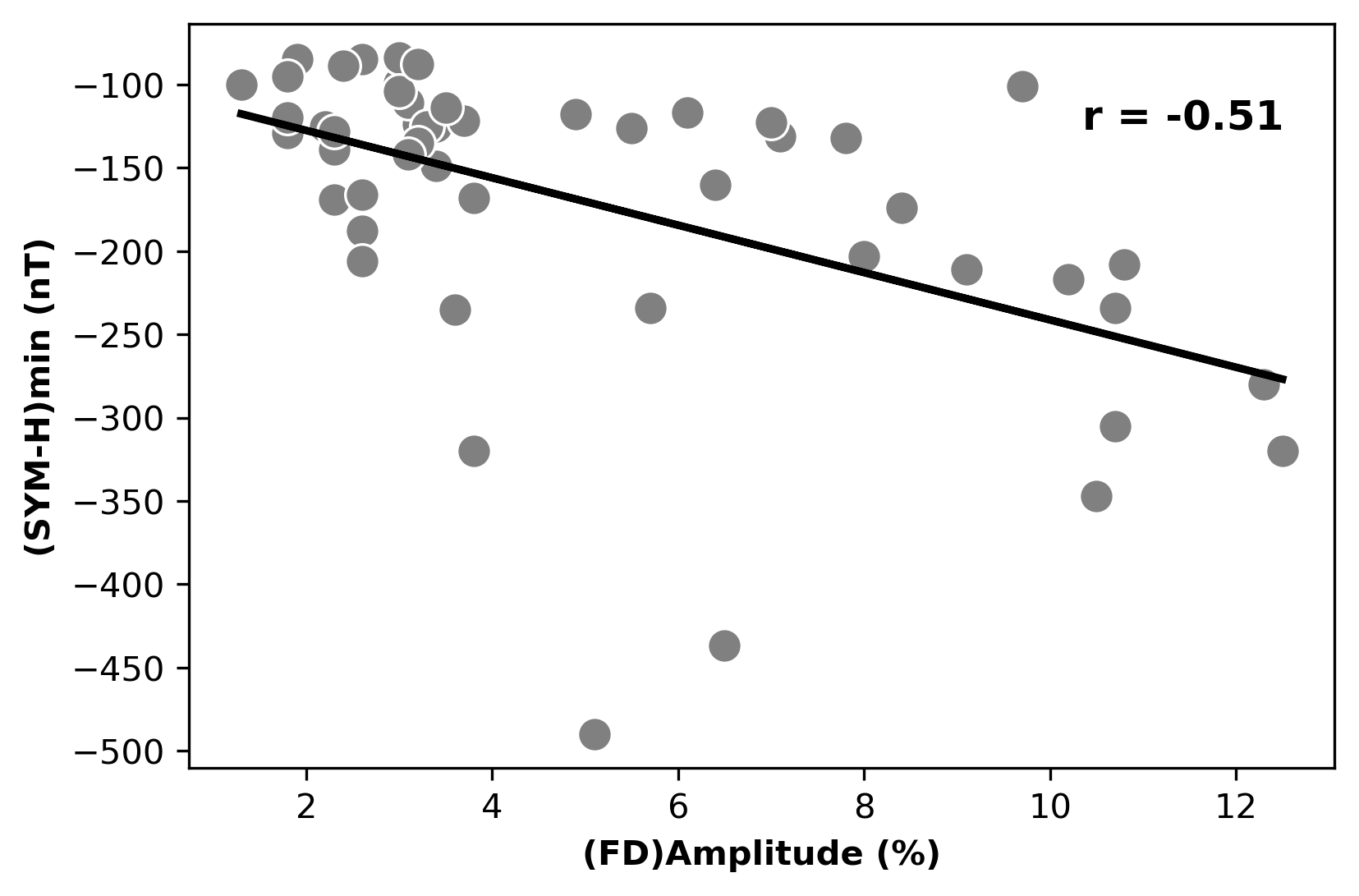}(a)
	\includegraphics[width=0.37\linewidth, height=4.3cm]{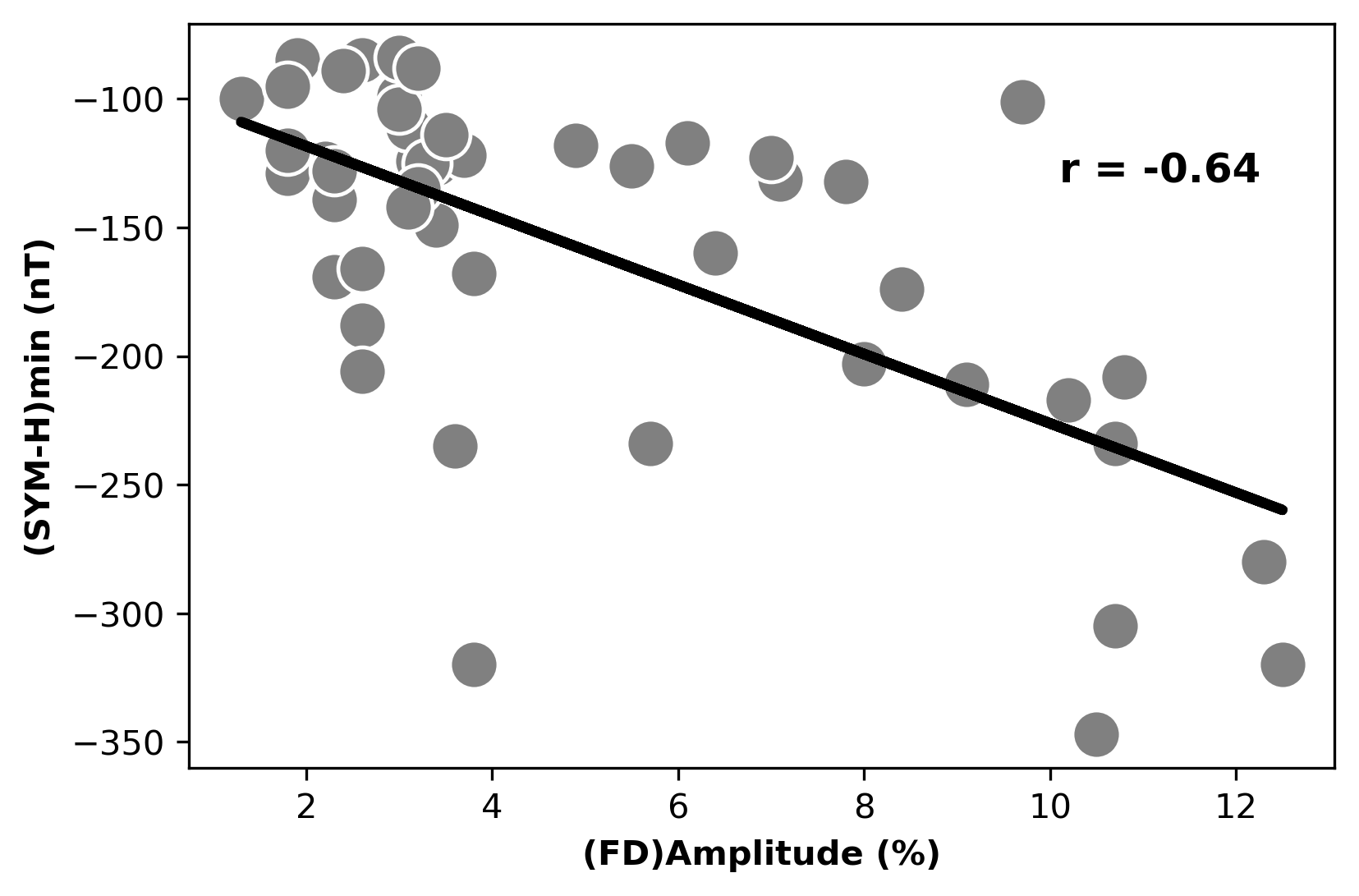}(b)
	\\[\smallskipamount]
	\includegraphics[width=0.37\linewidth, height=4.3cm]{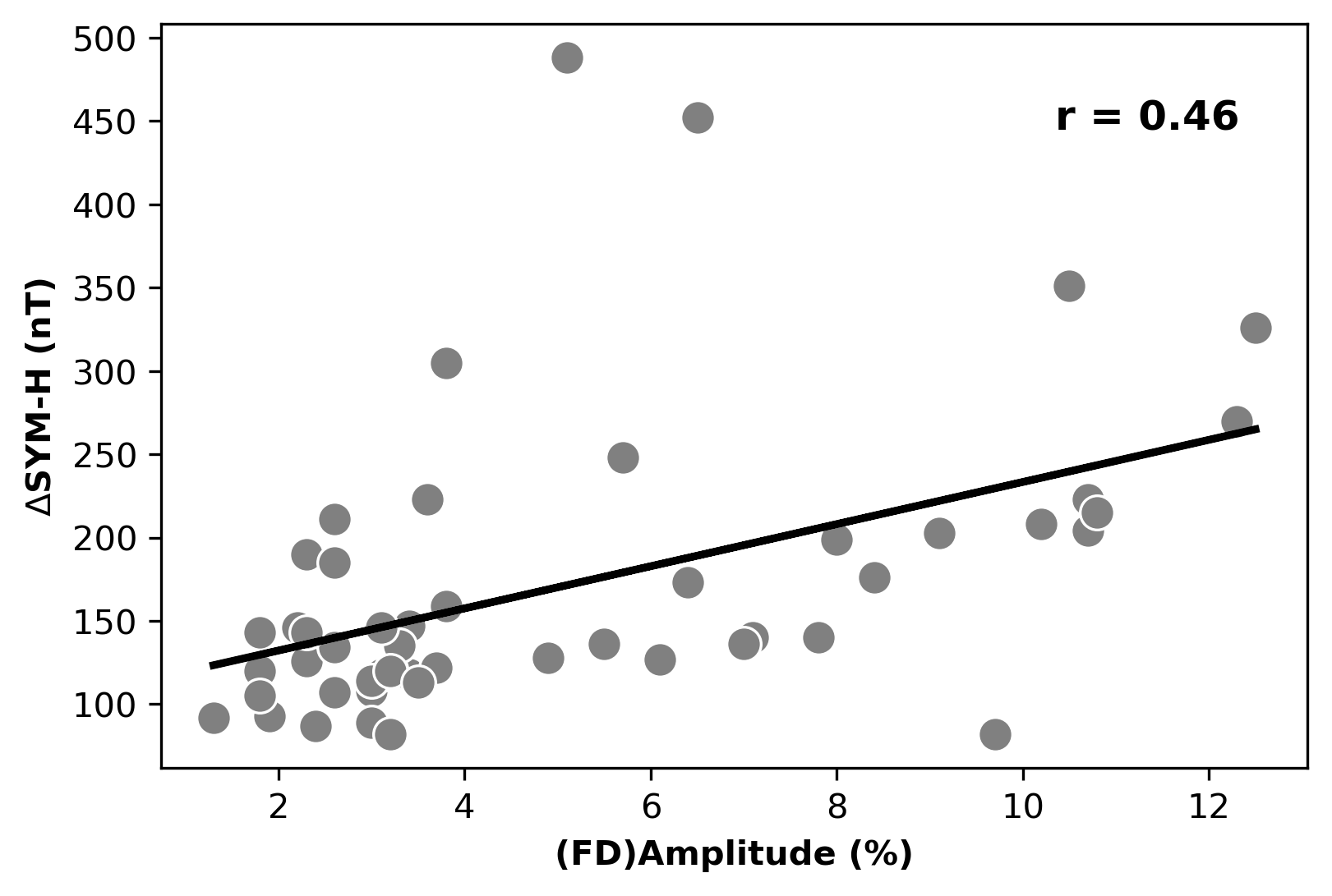}(c)
	\includegraphics[width=0.37\linewidth, height=4.3cm]{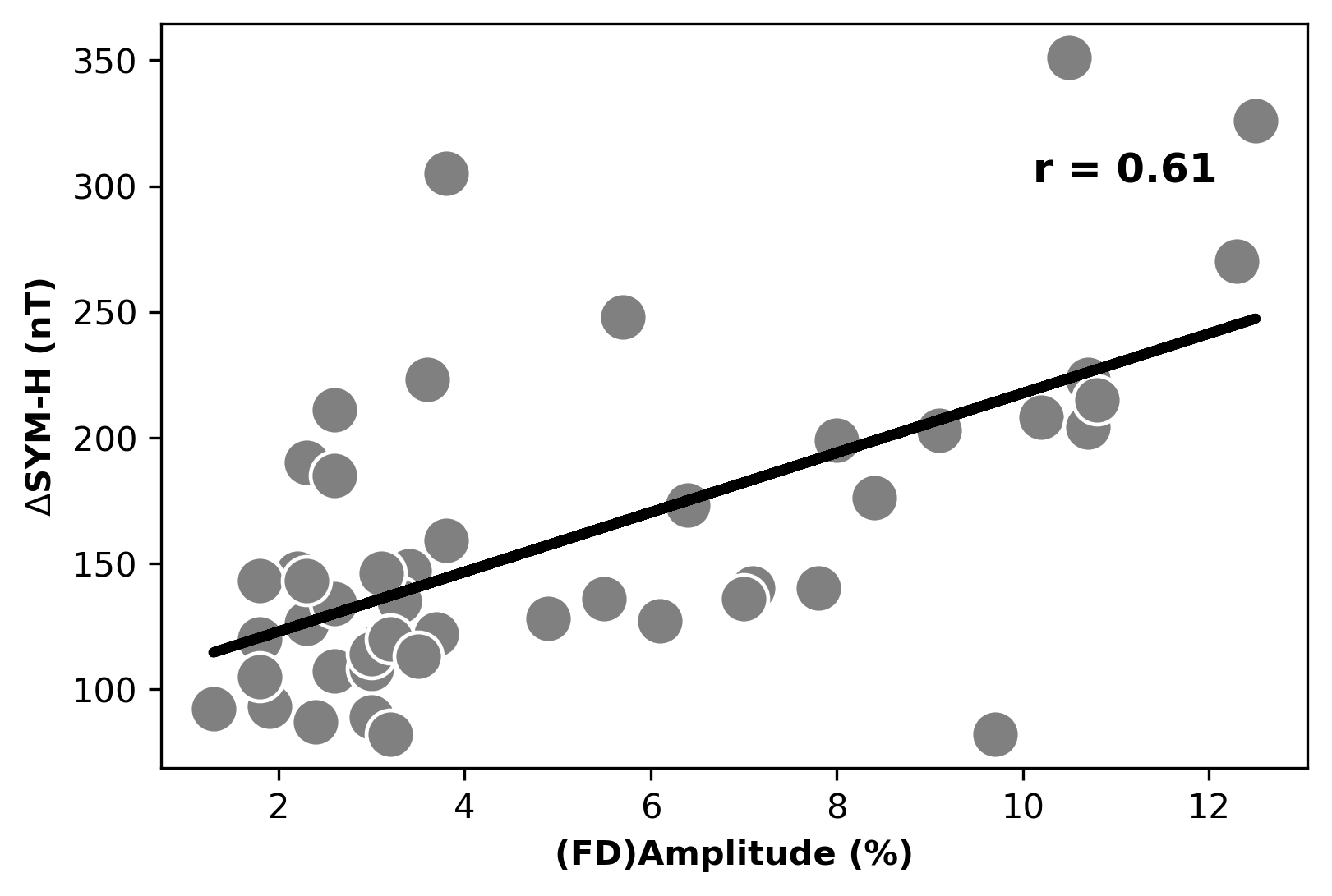}(d)
	\caption{Scatter plot of the correlation between FD amplitude (\%) and SYM-H minimum (nT) in subplots (a) and (b), as well as the correlation between FD amplitudes (\%) and $\Delta$SYM-H (nT) in subplots (c) and (d) for CME events. The right column subplots (b) and (d) represent moderate and strong storms, excluding extreme events. The grey-filled circles indicate the data points, while the black lines depict the linear fit.}\label{fig:2}
\end{figure*}

We selected CR intensity, various IP parameters and their products based on the FD amplitude.
We employed a linear fit function, $\mathrm{y(x)=ax+b}$, to model the relationship between the peak values of numerous IP parameters and the amplitude of FDs, in order to identify the best parameter that describes the main phase of FDs. Searching the best IP plasma and field parameters is the most crucial part of space weather.
Derived parameters have been used to effectively represent the main phase of geomagnetic disturbances \citep{1975JGR....80.4204B, 1981SSRv...28..121A, 2007JGRA..112.1206N, 2021AdSpR..68.4702B, 2023SpWea..2103314M, Ahmed2024a}. Similarly, these parameters have also been applied to describe the main phase of FDs \citep[e.g.,][]{2017SoPh..292..135A, 2019AdSpR..63.1100M}.
For this purpose, we have derived various coupling functions and tested their relationship with the FD amplitude.\\
We divided the IP parameters into three categories: one parameter, two and three parameter derivatives. To this purpose, IMF (total B, north-south Bz, perturbed $\sigma$B), the ratio of plasma pressure ($\beta$), ratio of perturbed to total IMF ($\sigma$B/B), plasma temperature (T), the plasma density  ($\rho$), the solar wind speed (v) were grouped as single parameters.
The dynamic pressure (P), dawn-dusk electric field (Ey), IP electric field (vB and B$\times$v) are derivatives of two IP parameters.
The parameters of $\rho$vB, $\sigma$B$\times$B$\times$v, ($\sigma$B/B$)\times$B$\times$v, P$\times$(vB) are derivatives of three parameter functions. We selected these combined functions to identify the optimal parameters that effectively explain the main phase of the FD. The solar wind speed (v) and the IMF magnitude (B) were found to exhibit a stronger correlation with the main phase FD, indicating their critical role in influencing the FD dynamics. These parameters are closely associated with the intensity and variability of the IP electric field, which governs the modulation of cosmic rays during solar-driven disturbances \citep[e.g.,][]{2021AdSpR..68.4702B} To refine the analysis further, we explored various combinations of these parameters, incorporating the plasma density ($\rho$) as a viscous term \citep[e.g.,][]{2007JGRA..112.1206N,Ahmed2024a,Ahmed2024}. This approach focused on developing a combined function that captures the interaction between these factors, providing a clearer and more accurate understanding of the processes influencing the main phase FD.
This classification is crucial for search the optimum parameter that can most accurately forecast the FD. This, in turn, serves as an important signature for predicting the magnetic storm's start time and time lag, as discussed in Section \ref{sub:lag}.\\
Figure \ref{fig:8} displays a scatter plot showing the relationship between the amplitude of FD and the peak values of the IP parameters. The best two parameters from each category (one parameter, two and three parameters) were displayed. The correlation of the remaining IP parameters with the amplitude of FD is summarized in Table \ref{taba}.
The data points (filled grey circles) are shown along with the best-fit linear relationship, indicated by a black line.
With respective Pearson's correlation coefficients of $\mathrm{r=0.73}$ and $\mathrm{r=0.58}$, the single parameters of plasma speed (v) and IMF total (B) have been found to best characterize the FD amplitude.
The FD amplitude has been best characterized by the two parameter derivatives of IP electric field related functions (B$_{\mathrm{max}}\times$v$_{\mathrm{max}}$) and [(vB)$_{\mathrm{max}}$], with Pearson's correlation coefficients of $\mathrm{r=0.68}$ and $\mathrm{r=0.65}$, respectively. The three-parameter derivative of the ratio of IMF perturbation to total magnetic field, merged with the IP electric field function [$\mathrm{(\sigma B/B)_{max}\times B_{max}\times v_{max}}$], and the dynamic pressure function, merged with the IP electric field [$\mathrm{P_{max}\times(vB)_{max}}$], best represent the FD amplitude. These correlations have Pearson's coefficients of $\mathrm{r=0.62}$ and $\mathrm{r=0.58}$, respectively.\\
Our result is in agreement with previous works \citep{2002Ap&SS.281..651B, 2009A&A...494.1107S, 2020ApJ...896..133L} who suggests that the degree of turbulence near the CME/ejecta may influence the extent of the FD.
Although there was no correlation observed between CR variability and the IMF north-south component Bz or proton density $\rho$, it has been demonstrated to be correlated with solar wind speed v \citep{2005JASTP..67..907K}, which is consistent with our findings.
\cite{2014JPhCS.511a2057V} pointed out that, the correlation between the speed of CMEs and the magnitudes of FDs associated to strong geomagnetic disturbances show a correlation coefficient of r $\sim0.70$. On the other hand, \cite{lingri2016forbush} discussed that, solar wind and CME velocities do not appear to be correlated with the amplitude of FDs.
\cite{2020ApJ...896..133L} found a substantial relationship between maximum plasma flow speed and FD amplitude, which is consistent with our findings. However, \cite{2023JASTP.24205981B} examined variations in solar wind parameters, such as velocity, plasma density, and the IMF-Bz component, along with the Earth's Dst index, in relation to CR flux data from eight neutron monitor stations. Through wavelet analysis of CR intensity, they identified a relationship between FD amplitude and plasma density, which stands in contrast to our findings.\\
Our focus is on potential coupling function searches that more accurately capture the main phase evolution of FD.
In this sense, the main phase properties of the FD are best described by the electric field related coupling function vB. The function vB was extensively studied previously \citep[e.g.,][]{1979GeoRL...6..577K, 1983JGR....88.5727W}. \cite{2021AdSpR..68.4702B} reported that the IP electric field function vB best describes the amplitude of FD, with a Pearson's correlation coefficient of $\mathrm{r\sim0.669}$. Our findings are consistent with theirs. In addition to function vB, they reported that plasma speed, with a Pearson's correlation coefficient of $\mathrm{r\sim0.636}$, is the second best parameter, which is somewhat less than our result of $\mathrm{r=0.73}$.
Overall, we propose that the three-parameter derivative of the ratio of IMF perturbation to the total magnetic field, combined with the IP electric field function [$\mathrm{(\sigma B/B)_{max}\times B_{max}\times v_{max}}$], provides the most comprehensive explanation for the main phase FD.
We compare our results to those reported by \cite{Ahmed2024a}, who examined the relationship between the amplitude of Dst ($\Delta$Dst) and IP parameters.
The IMF (total B and north-south component Bz) best represents the amplitude of Dst, with Pearson's correlation coefficients of $0.85$ and $-0.84$, respectively.
Comparing our result of FD amplitude with the geomagnetic disturbance amplitude of \cite{Ahmed2024a}, we found that despite differences in the number of events investigated, geomagnetic disturbances exhibited a better association with IP parameters during their main phase peak values than FD amplitudes. There are comparatively strong relationships between FDs and geomagnetic disturbances for the total IMF B.
The statistical significance of the correlations between FD amplitude and the IP parameters was determined by calculating p-values, as illustrated in Figure \ref{fig:81} and summarized in Table Table \ref{taba}.\\
\begin{table*}[h]
	\centering
	\caption{The correlation constants in Figure \ref{fig:8} for six parameters, along with the remaining parameters.}\label{taba}
	\begin{tabular}{@{}lccccccc@{}}
		\toprule
		parameter & r & slope&p-value & parameter & r& slope&p-value \\
		\midrule
		B$_{\mathrm{max}}$ & 0.58& 0.14 &2.8e-4 & (Ey)$_{\mathrm{max}}$ & 0.41&0.13&1.4e-2 \\
		(Bz)$_{\mathrm{min}}$ & -0.35 &-0.10&3.7e-2 & $\mathrm{\beta_{min}}$&-0.29& -4.86&9.4e-2 \\
		($\sigma$B)$_{\mathrm{max}}$ & 0.55 &0.41&7.0e-4 & (vB)$_{\mathrm{max}}$ & 0.65&0.16& 2.0e-5 \\
		($\sigma$B/B)$_{\mathrm{max}}$ & 0.21 &4.56&2.3e-1 & ($\rho$vB)$_{\mathrm{max}}$ & 0.55&0.01&6.0e-4  \\
		T$_{\mathrm{max}}$ & 0.55 &0.00&5.9e-4 & $\mathrm{B_{max}\times v_{max}}$ & 0.68& 1.55& 5.8e-6 \\
		$\mathrm{\rho_{max}}$ & 0.03 &0.01&8.7e-1 & ($\sigma$B)$_{\mathrm{max}}\times\mathrm{B_{max}\times v_{max}}$ & 0.59& 0.79&1.6e-4 \\
		v$_{\mathrm{max}}$ & 0.73 &0.01&6.3e-7 & ($\sigma$B/B)$_{\mathrm{max}}\times\mathrm{B_{max}\times v_{max}}$ & 0.62& 2.13& 6.6e-5 \\
		P$_{\mathrm{max}}$ & 0.54 & 0.11&8.9e-4 & P$_{\mathrm{max}}\times$(vB)$_{\mathrm{max}}$ & 0.58& 0.26&3.0e-4 \\
		\botrule
		\multicolumn{8}{l}{\footnotesize r is the Pearson's linear correlation coefficient.}
	\end{tabular}
\end{table*}
Figure \ref{fig:2} illustrates the relationship between the amplitude of the CR decrease, represented by FD, and the peak values of geomagnetic storm, as shown in Figure \ref{fig:2} first-row (a \& b) for CME-triggered events. Additionally, the figure presents the correlation between the amplitudes of FD and geomagnetic disturbance, indicated by $\Delta$SYM-H (Figure \ref{fig:2}, c \& d) for the same events.
The amplitude of SYM-H is determined by subtracting the SYM-H value at the peak time from the SYM-H value at the onset time, expressed as:
\begin{equation}
\mathrm{\Delta SYM-H=(SYM-H)_{onset}-(SYM-H)_{peak}}\label{eq3}
\end{equation}
The black line displays the linear fit, while the grey filled circles represent the data points.
The first-column subplots show the correlation between FD amplitude with minimum and amplitude of SYM-H with a Pearson correlation coefficients of $-0.51$ and $0.46$, respectively.
Excluding severe events ($\mathrm{SYM-H}\leq-350$ nT)\footnote{The most often used geomagnetic index, Dst, was used to classify the disturbance: a weak storm has $-30$ nT to $-50$ nT, a moderate storm has $-50$ nT to $-100$ nT, a strong storm has $-100$ nT to $-200$ nT, a severe storm has $-200$ nT to $-350$ nT and a great storm has Dst $< 350$ nT \citep[e.g.,][]{1997JGR...10214209L}.} improved the fit to $-0.64$ and $0.61$ as indicated by second-column subplots, respectively.
The FD amplitude and the peak of SYM-H exhibit a correlation, though the strength of this correlation is moderate.
Unlike the results of \cite{2005JASTP..67..907K, 2010AnGeo..28..479K}, who reported no clear proportional relationship between the magnitudes of FD and geomagnetic storms, our study reveals a moderate correlation.
This can be the result of different mechanisms causing the two occurrences \citep[e.g.,][]{MUSTAJAB201343}.
These findings suggest a better correlation between FD amplitude with the peak and amplitude values of geomagnetic storms during moderate and strong magnetic storms \citep{2016SoPh..291.1025L} compared to extreme CME-driven storms. In the case of extreme CME-driven storms, the geomagnetic response is often complicated by the presence of multiple, successive CME events and prolonged periods of southward IMF Bz, which can lead to saturation effects in the magnetospheric response. These conditions may cause the FD amplitude to decouple from a simple linear relationship with peak storm values due to nonlinear geomagnetic responses, shielding effects, or multiple-phase storm developments. In contrast, during moderate and strong single-event storms, the magnetospheric response tends to be more direct and predictable, leading to a better correlation between FD amplitude and storm parameters.
\cite{2024Ge&Ae..64..289B} pointed out that the FD magnitude shows a nonlinear dependence on geomagnetic storm intensity, with moderate correlation to geomagnetic indices (Ap, Kp, Dst). However, the timing of peak CR and geomagnetic responses varies and is influenced by the Bz-component of the IMF.
The number of CIR-driven events was statistically insufficient to undertake correlation analysis, so we omitted it.
\subsection{FD correlation with ICME transient speed and duration}\label{transient}
\begin{figure*}[t]
	\centering
	\includegraphics[width=0.84\linewidth, height=3.8cm]{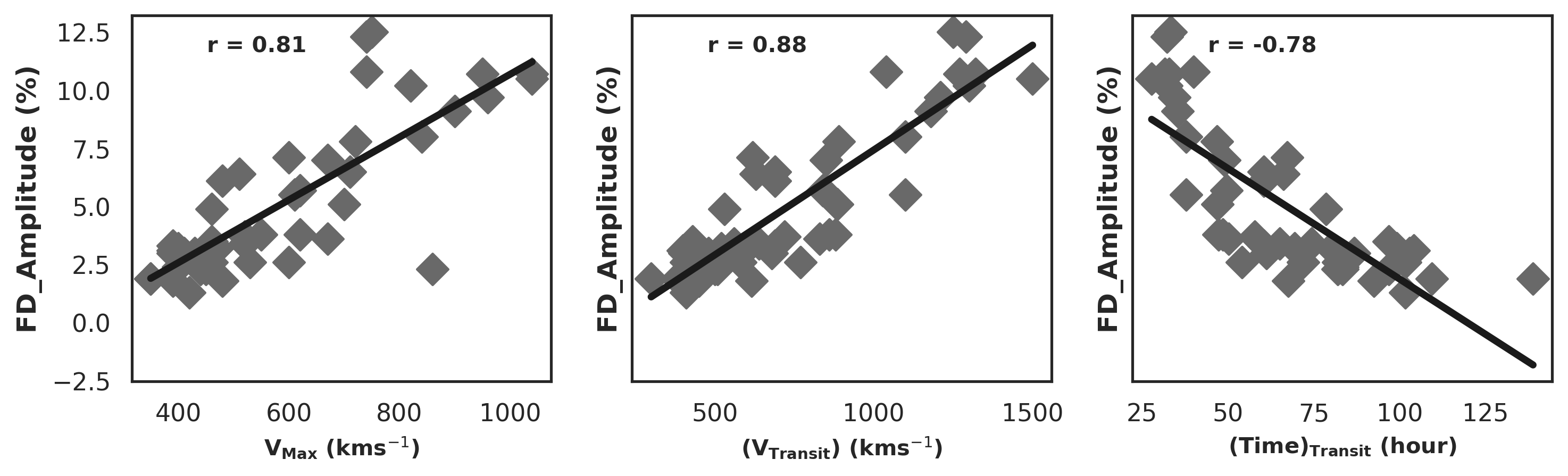}(a)
	\includegraphics[width=0.84\linewidth, height=3.8cm]{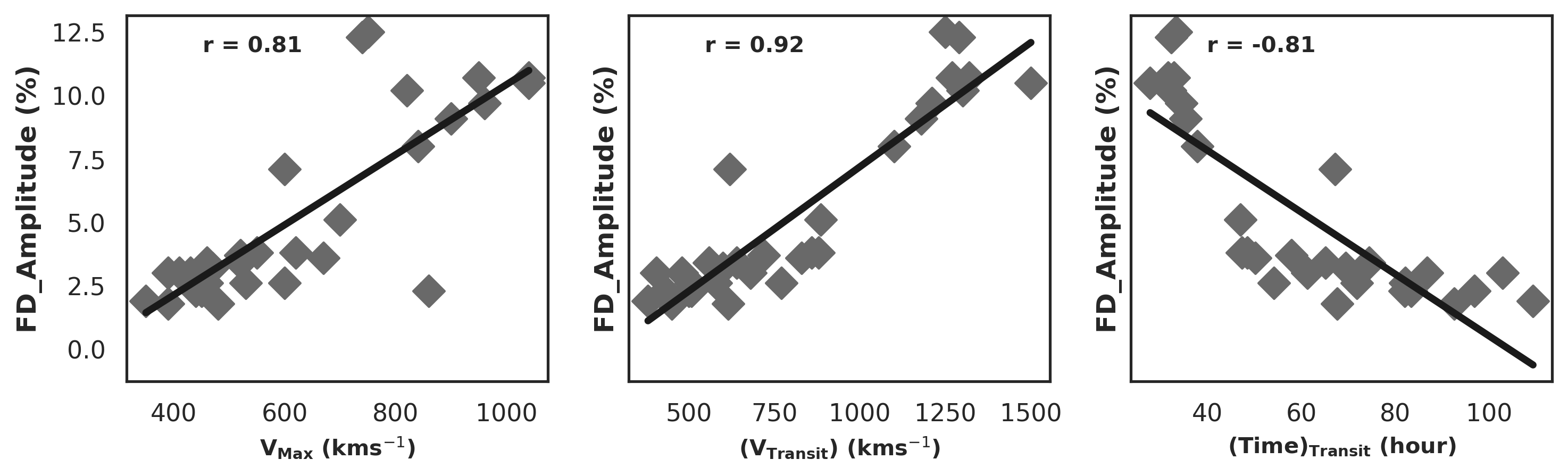}(b)
	\includegraphics[width=0.84\linewidth, height=3.8cm]{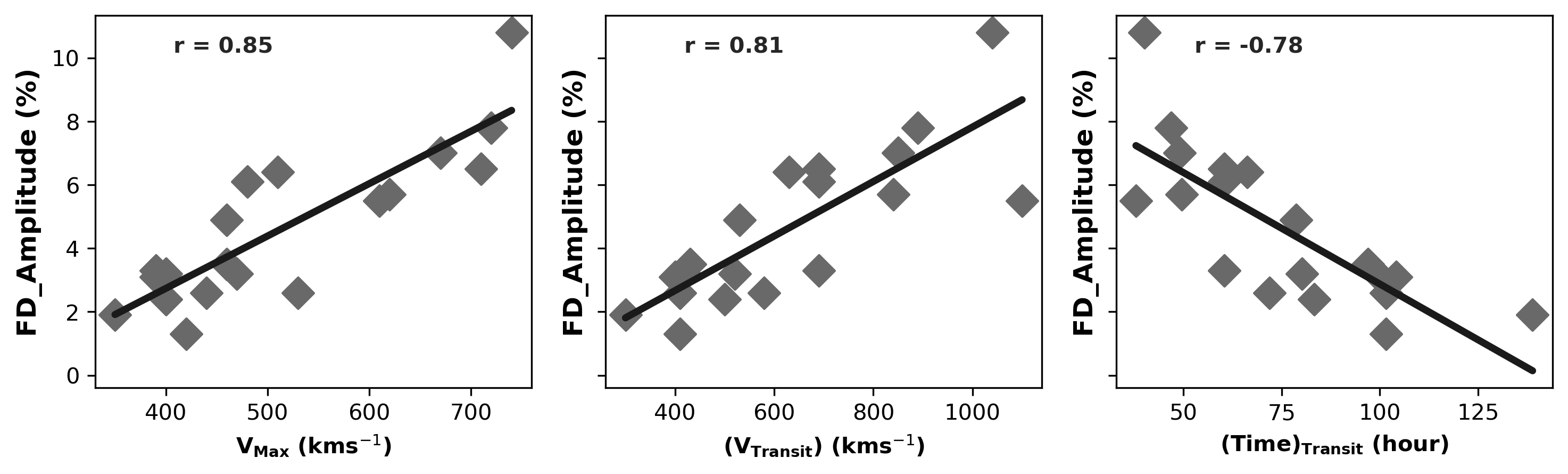}(c)
	\caption{Scatter plot of the correlation between FD amplitude and ICME (maximum speed, transit speed and transit time). The gray diamonds represent the data points, while the black lines correspond to the best-fit. The first row, labeled as (a), illustrates ICME with both SH+non-SH. The second row, labeled as (b), depicts ICME with SH, while the third row, labeled as (c), represents ICME without SH.}
	\label{fig:13}
\end{figure*}

Figure \ref{fig:13} illustrates the scatter plot which shows the relationship between FD amplitude and ICME (the CME in the IP space) as the ejection from the sun's surface propagates through the IP space till it reaches 1 AU at the Earth's orbit.
The ICME transient speed and time were derived from the near-Earth ICME table\footnote{\url{https://izw1.caltech.edu/ACE/ASC/DATA/level3/icmetable2.htm}} \citep{2003JGRA..108.1156C, Richardson2024}, while the ICME's angular width during ejection was taken from the SOHO/LASCO CME catalog\footnote{\url{https://cdaw.gsfc.nasa.gov/CME_list/}}. Forty-eight events with FD amplitude ranging from 1.3 -- 12.5\%, have been analyzed.\\
The IP parameters of ICME: solar wind speed upstream disturbances (dv), mean speed ($\mathrm{v_{ICME}}$), maximum speed ($\mathrm{v_{max}}$), average speed of from ejection to Earth's orbit ($\mathrm{v_{transit}}$), duration from ejection to Earth's orbit ($\mathrm{transit_{time}}$), beginning of disturbance to onset of FD taken as sheath period ($\mathrm{SH_{time}}$), main phase period ($\mathrm{ICME_{time}}$), duration from beginning of disturbance to peak of FD ($\mathrm{(SH+ICME)_{time}}$), and the ejection angular extent (angular width) have been correlated with FD amplitude. The units of measurement for all velocities are kms$^{-1}$, periods are hours, and degrees of angle width.
We computed the ICME transit time using the period between the ejection time (derived from SOHO LASCO) and the commencement of FD. Earlier studies by \citep[e.g.,][]{2005JGRA..110.9S05P} reported that anti-correlation between FD recovery time and transit speed of the ICME, even though our analysis was based on main phase FD.\\
We categorized the data into three groups based on the presence of a forward shock ahead of the ICME: SH+non-SH (fast forward shock ahead of the compression sheath region + without shock), SH (fast forward shock ahead of the compression sheath region), and non-SH (ICME without shock). Out of the 51 selected ICME events, 36 had fast forward shock ahead of the compression sheath region, 12 lacked shocks, and 3 had data gaps.
A linear correlation was used to effectively represent the relationship between FD amplitude and the ICME transit speed, duration, and angular extent for the three ICME event groups. For visibility purpose, we displayed the correlations of three ICME parameters: transit speed, transit period, and angular extent from each group, as shown in Figure \ref{fig:13}. The results for the remaining parameters from each group are provided in Table \ref{tabb}.
The first row (a), shows forward shock ahead of the compression sheath region and ICME without shock/sheath. The second row (b) exhibits fast forward shock ahead of the compression sheath region, while the third row (c) shows ICME without shock/sheath.
The gray-filled diamonds represent the data points, while the black lines indicate the best-fit lines.\\
The transit velocity exhibits a stronger correlation with the FD amplitude, with correlation coefficients of $0.88$, $0.92$, and $0.81$ for SH+non-SH, SH, and non-SH events, respectively.
This finding contradicts the results of previous studies, such as \cite{lingri2016forbush}, which reported no significant correlation between FD amplitude and the velocities of CMEs.
The differences may primarily result from the variation in cutoff rigidity. They used 10 GV, whereas our study used 0.81 GV. Additionally, the number of events analyzed could be a contributing factor, as their analysis included only 15 CME-driven events.
Among the three groups, the relationship between FD amplitude and ICME parameters for events with a fast forward shock ahead of the compression sheath region shows a strong correlation, as illustrated in Figure \ref{fig:13}(b).
However, ICME events without a shock/sheath, as shown in Figure \ref{fig:13}(c), exhibit a relatively weaker correlation compared to ICME events with a shock.
High speed ICMEs generate higher FD amplitudes, while low speed transients yield lower amplitudes. From the selected fifty one FD events $\sim71\%$ shows fast forward shock ahead of the compression sheath region, $\sim23\%$ without shock and $\sim6\%$ with data gap as taken from ACE and WIND observations. Those ICME events with data gaps have been excluded from the analysis.\\
\cite{2014SoPh..289.3949B} discussed the correlation between FD amplitude and the speed of ICME. They reported that, the most significant FDs are caused by large CMEs, which are defined by their high velocity and broad angular width.
We analyzed the relationship between angular extent and various ICME parameters. Our results indicate that angular extent did not demonstrate any correlation with ICME speed or duration.
\begin{table*}[h]
	\caption{The correlation constants of Figure \ref{fig:13} for three groups (SH+non-SH, SH \& non-SH), along with the remaining parameters.}\label{tabb}%
	\centering
	\begin{tabular}{@{}lccccccccc@{}}
		\toprule
		ICME parameters&\multicolumn{3}{c}{SH$+$non-SH}&\multicolumn{3}{c}{SH}&\multicolumn{3}{c}{non-SH}\\
		&r&s&p-value&r&s&p-value&r&s&p-value\\
		\midrule
(dv)$_\mathrm{shock}$ speed & 0.81& 0.02&2.03e-12 &0.83& 0.02&1.75e-8 &0.85&0.03&4.43e-6 \\
		(ICME)$_\mathrm{mean}$ speed & 0.77&0.02 &1.11e-10 & 0.78& 0.02&3.72e-7 &0.76&0.02&1.50e-4 \\
		(ICME)$_\mathrm{max}$ speed & 0.81&0.01 & 4.00e-12&0.81&0.01 &6.60e-8 &0.85&0.02&3.13e-6 \\
		(ICME)$_\mathrm{Transit}$ speed & 0.88&0.01&1.05e-16 & 0.92&0.01&8.23e-13 & 0.81&0.01&2.45e-5 \\
		(ICME)$_\mathrm{Transit}$ time & -0.78& -0.10&9.23e-11 &-0.81&-0.12 &5.20e-8 &-0.78& -0.07&8.87e-5 \\
		Sheath time & 0.01& 0.01&9.23e-1 &-0.07&-0.07 &7.12e-1 &0.21&0.08&3.85e-1 \\
		ICME time & 0.34& 0.06&1.80e-2 &0.35&0.07 &5.82e-2 &0.27& 0.06&2.68e-1 \\
		(SH+ICME) time & 0.33&0.06&2.12e-2 & 0.33&0.06 & 7.21e-2&0.32&0.06&1.75e-1 \\
		Angular width & 0.18&0.00&2.22e-1 & 0.13& 0.00&5.08e-1 &0.32&0.00&1.84e-1 \\
		\botrule
		\multicolumn{10}{l}{\footnotesize SH stands for forward shock.}\\
		\multicolumn{10}{l}{\footnotesize r is the Pearson's linear correlation coefficient.}\\
		\multicolumn{10}{l}{\footnotesize s stands for slope of the equation.}\\
	\end{tabular}
\end{table*}
\subsection{Cutoff rigidity and energy dependence of CR}\label{cutoff}

 \begin{sidewaystable*}
\footnotesize
\caption{List of neutron monitors along with their cutoff rigidity, energy spectrum, geographic altitude and latitude profiles. The start date and amplitudes of FD for the selected 9 events as well as parameters for the analysis from Figure \ref{fig:9} and Figure \ref{fig:10} were tabulated. \label{table01}}
\begin{tabular*}{\textheight}{@{\extracolsep\fill}lccccccccccccc}
\toprule%
 && \multicolumn{12}{@{}c@{}}{neutron monitors}\\ \cmidrule{3-14}
No&Start Date & Thule	& Oulu & Newark & Moscow & Jung$^\beta$ & Hermanus&Rome&Mexico&Tsumeb&Halekala&Tibet&Sir$^\alpha$\\
          & & 0.30$^a$ & 0.81$^a$ & 2.40$^a$ & 2.43$^a$ & 2.49$^a$ & 4.58$^a$&6.32$^a$&8.28$^a$&9.15$^a$&12.92$^a$&14.10$^a$&16.80$^a$ \\
       &    & 10.17$^b$ & 10.30$^b$ & 10.99$^b$ & 11.01$^b$ & 12.58$^b$ & 12.67$^b$ & 14.60$^b$ &17.41$^b$&18.87$^b$&26.75$^b$&29.73$^b$&37.46$^b$\\
          & & 26$^c$ & 15$^c$ & 50$^c$ & 200$^c$ & 3570$^c$ & 26$^c$ & 0$^c$ & 2274$^c$ &1240$^c$&3030$^c$&0$^c$&2565$^c$\\
          & & 76.5N$^d$ & 65.05N$^d$ & 39.68N$^d$ & 55.47N$^d$ & 46.55N$^d$ & 34.42S$^d$ & 41.86N$^d$ &19.33N$^d$&19.20S$^d$&20.72N$^d$&30.11N$^d$&18.59N$^d$\\
         &  & 68.70W$^e$ & 25.47E$^e$ & 75.75W$^e$ & 37.32E$^e$ & 7.98E$^e$ & 19.23E$^e$ & 12.47E$^e$ &260.82E$^e$&17.58E$^e$&156.28W$^e$&90.53E$^e$&98.49E$^e$\\
 \midrule
 i&1998-08-26T11:00 & 7.5 & 7.6 & 7.6 & 7.4 & 7 & 6.3 &5.8&4.4&4.5&4.4&-&-\\
 ii&1998-09-24T19:00 & 9.2 & 10 & 9.7 & 9.6 & 9 & 8.3 &7.5&5.3&6&4.6&-&-\\
 iii&2000-07-15T12:00 & 16.4 & 12.3 & 16.2 & 14.8 & 13.1 & 13 &10.6&11&9.6&9.4&7.8&-\\
 iv&2001-04-11T17:00 & 11 & 12.3 & 11.5 & 11.8 & 10.9 & 9.7 &9&8.6&7.7&6.3&5.5&-\\
 v&2001-11-05T19:00 & 10.7 & 12.5 & 9.4 & 10.9 & 8.4 & 8 &6.9&6&5.2&5.8&5.6&-\\
 vi&2001-11-24T06:00 & 8.9 & 10.7 & 8.8 & 9.8 & 8.5 & 7 &5.8&5&5.6&5.5&4&-\\
 vii&2005-01-21T18:00 & 8.8 & 8.6 & 9.4 & 9.2 & 6 & 6.4 &5.9&5&5&4.5&3.4&-\\
 viii&2005-05-15T02:00 & 10.1 & 10.7  & 10.5 & 9.3 & 7.1 & 6.8 &6.2&6.5&-&6&6.5&-\\
 ix&2015-06-22T03:00 & 10 & 10.8 & 9.5 & 10 & 9.3 & 7.6 &7&6.5&6.8&-&-&5.4\\
 \toprule%
 && \multicolumn{3}{@{}c@{}}{cutoff rigidity}&\multicolumn{3}{@{}c@{}}{energy spectrum}&SYM-H$_{\mathrm{minimum}}$&&&&&\\ \cmidrule{3-5}\cmidrule{6-8}
 && slope	& r & p-value & slope & r & p-value&(nT)&&&&&\\
  \midrule
 i&1998-08-26T11:00 & -0.32   &   -0.94& 5.79e-5 & -0.23 & -0.88& 8.20e-4 &-174&&&&&\\
 ii&1998-09-24T19:00 & -0.47   &   -0.95& 2.68e-5 & -0.35 & -0.93& 1.19e-4 &-217&&&&&\\
 iii&2000-07-15T12:00 & -0.61   &   -0.95& 5.75e-6 & -0.39 & -0.89& 2.77e-4 &-347&&&&&\\
 iv&2001-04-11T17:00 & -0.46   &  -0.97& 6.37e-7 & -0.32 & -0.96& 3.06e-6 &-280&&&&&\\
 v&2001-11-05T19:00 & -0.44   &  -0.91& 1.07e-4 & -0.27 & -0.80& 2.97e-3 &-320&&&&&\\
 vi&2001-11-24T06:00 & -0.42   &  -0.90& 1.41e-4 & -0.27 & -0.83& 1.68e-3 &-234&&&&&\\
 vii&2005-01-21T18:00 & -0.41   &   -0.92& 5.26e-5 & -0.27 & -0.85& 8.35e-4 &-101&&&&&\\
 viii&2005-05-15T02:00 & -0.38   &   -0.88& 8.73e-4 & -0.23 & -0.78& 7.98e-3 &-305&&&&&\\
 ix&2015-06-22T03:00 & -0.34   &  -0.91& 2.82e-4 & -0.17 & -0.79& 6.98e-3 &-208&&&&&\\
 \botrule
 \end{tabular*}
\footnotetext{$^a$ -- Cutoff rigidity (GV), s -- stands for slope}
\footnotetext{$^b$ -- Median energy (GeV), r -- is Pearson's correlation coefficient}
\footnotetext{$^c$ -- Altitude (m), p -- is p-value of the statistical significance value}
\footnotetext{$^d$ -- Geographic latitude}
\footnotetext{$^e$ -- Geographic longitude}
\footnotetext{$^\beta$ -- Jungfraujoch neutron monitor}
\footnotetext{$^\alpha$ -- Princess Sirindhorn neutron monitor}
\end{sidewaystable*}

\begin{figure*}[h]
	\centering
	\includegraphics[width=0.75\linewidth]{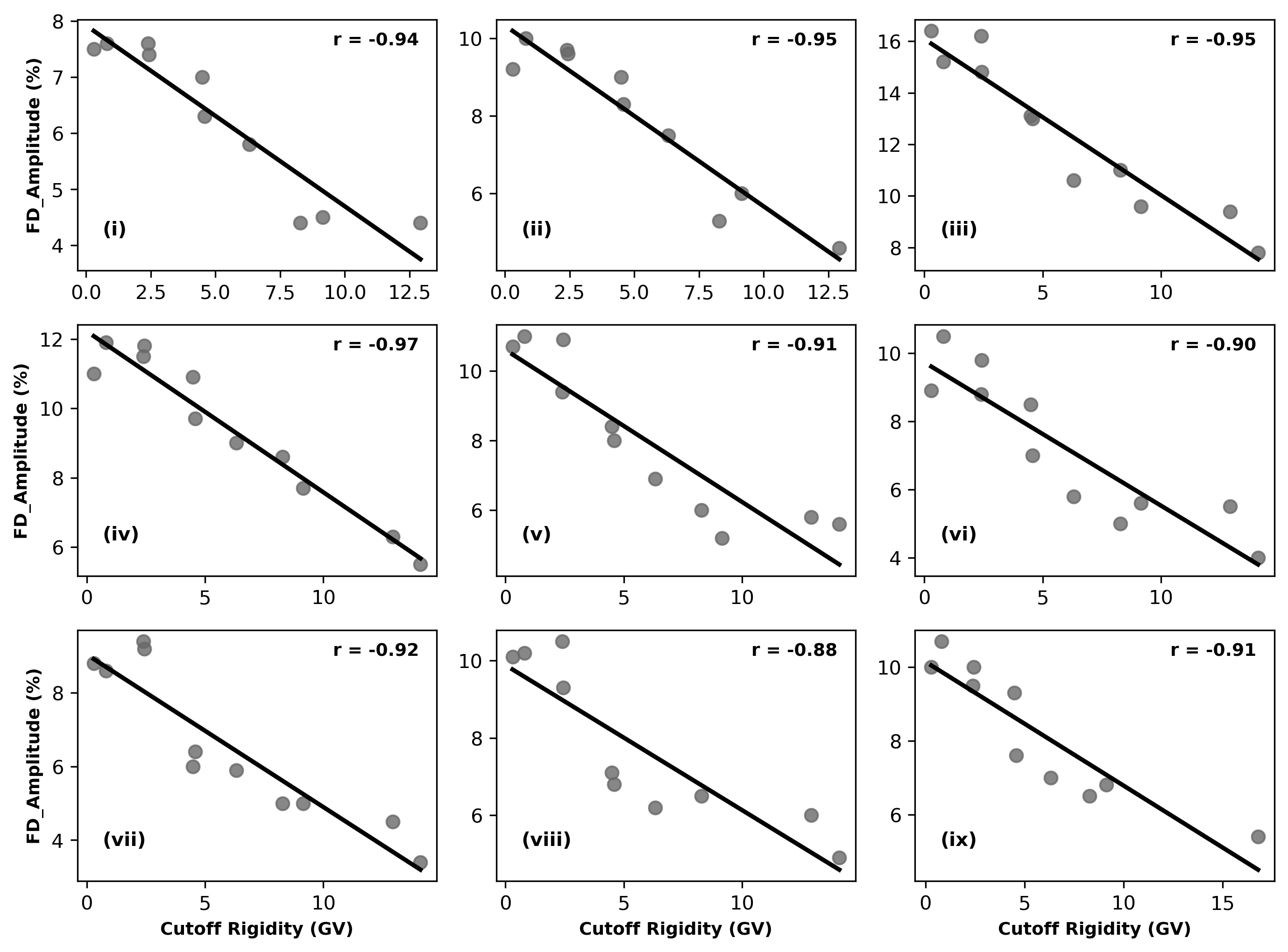}
\caption{Scatter plot shows the relationship between FD amplitude and cutoff rigidity for the selected 9 good events. The grey filled circles are the data points and the black lines are the linear best fit.	\label{fig:9}}
\end{figure*}
\begin{figure*}[h]
	\centering
	\includegraphics[width=0.75\linewidth]{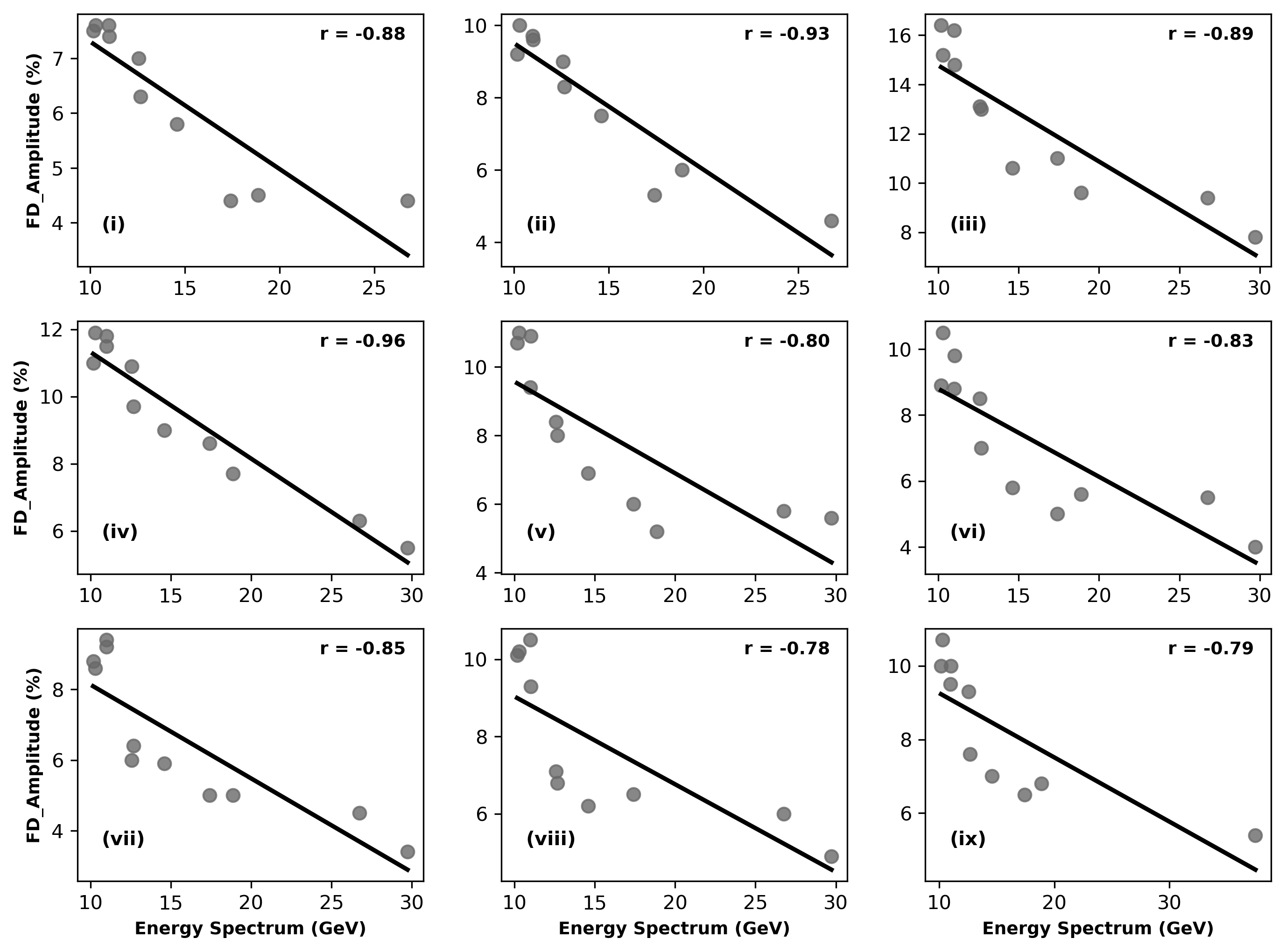}
	\caption{Scatter plot the same as Figure \ref{fig:9} except energy spectrum.}
	\label{fig:10}
\end{figure*}

The general characteristics of the temporal variation in the rigidity spectrum of FDs were investigated \citep[e.g.,][]{2013SoPh..286..561A}. Geomagnetic storms significantly disturb the magnetic cutoff rigidity, leading to distortions in the time profile of the FD in CR intensity. Neutron monitor data are affected by geomagnetic disturbances via changes in cutoff rigidity, which likely distort the rigidity spectrum of FDs \citep[e.g.,][]{2013JPhCS.409a2184A}. The apparent enhancement in CR flux during a FD, as detected by neutron monitors, may result from magnetospheric disturbances caused by geomagnetic storms \citep[e.g.,][]{2023JASTP.25206146G}.
The relationship between FDs and geomagnetic cutoff rigidity has been investigated \citep[e.g.,][]{nwuzor2024investigating}.\\
Table \ref{table01} summarizes the profiles of 12 ground-based neutron monitors from the World Neutron Monitor Network\footnote{\url{https://www.nmdb.eu/nest/}}, 
spanning a range of latitudes and altitudes with geomagnetic cutoff rigidities ranging from 0.3 to 17 GV  and energy spectrum from 10 to 37 GeV. The energy spectrum for these stations was derived from geomagnetic cutoff rigidity, following methods described in \citep[e.g.,][]{2008JGRA..113.7102U}.\\
Figure \ref{fig:9} and Figure \ref{fig:10} present scatter plots illustrating the relationship between FD amplitude with geomagnetic cutoff rigidity and energy for nine events with FD amplitudes exceeding 8.4\%, observed across these 12 neutron monitor stations\footnote{As these 12 stations listed in Table \ref{table01}}. 
Grey filled circles represent individual data points, and a black line indicates the linear fit. A linear correlation is observed between FD amplitude with geomagnetic cutoff rigidity and energy spectrum.\\
For the event beginning at 16:50 UT on 11 April 2001 (see Tables \ref{table01} and \ref{t1}), FD amplitude shows strong linear correlations with geomagnetic cutoff rigidity (Pearson coefficient $=-0.97$, Figure \ref{fig:9}) and with energy spectrum (Pearson coefficient $=-0.96$, Figure \ref{fig:10}). This event, associated with a high FD amplitude of 12.3\% generated a severe geomagnetic storm (SYM-H$_{\mathrm{minimum}}$ of $-280$ nT).
Constants derived from this analysis are listed in Table \ref{table01}.\\
FD amplitude typically follows a power-law dependence on geomagnetic rigidity \citep{2012AdSpR..50..725A, SAVIC20191483} and \citep[P$^\gamma$, with $\gamma$ ranging from 0.4 to 1.2;][]{2000SSRv...93...55C} confirms the same model.
FD amplitude is often reported to follow a power-law dependence on geomagnetic rigidity \citep{2012AdSpR..50..725A, SAVIC20191483}, with P$^\gamma$ \citep[$\gamma$ ranging from 0.4 to 1.2;][]{2000SSRv...93...55C} supporting the same model.
\cite{2002JGRA..107.1174L} showed that FD amplitude exhibits exponential decay during the main phase over a broad energy spectrum (0.6--50 GV).
During the main phase, FD amplitude exhibits exponential decay across a broad energy spectrum \citep[0.6--50 GV;][]{2002JGRA..107.1174L}. Our study reveals a two-step linear dependence of FD amplitude on energy and geomagnetic cutoff rigidity. In the lower rigidity range (0.3--7.5 GV) and energy spectrum (10.17--18.5 GeV), FD amplitude decreases sharply, while in higher rigidity regimes, the decrease is more gradual.
This behavior likely results from at low geomagnetic cutoff energy spectrum (0.3--17 GV) the FD amplitude correlated linearly while at high energy the interaction becomes more complex and reveals non-linear behavior.\\
In contrast, \cite{2016SoPh..291.1025L} reported a latitudinal dependence of FD amplitude, increasing proportionally with station latitude, and a nonlinear energy dependence based on three events (18 February 2011, 8 March 2012, and 14 July 2012; see their Figure 5). These findings differ from our linear dependence, possibly due to differences in the energy spectrum analyzed. Other studies, such as \cite{2013A&A...555A.139A} and \cite{2021ApJ...920L..43A}, support an exponential dependence of FD amplitude on energy. Additionally, FD amplitude peaks at low rigidities due to a higher number of blocked particle trajectories, but decreases sharply at high rigidities due to less restrictive geometric conditions \citep[e.g.,][]{2019ApJ...880...17P}. \\
Our two-step linear dependence aligns with recent work \citep[e.g.,][]{2024IAUGA..32P2411S}, suggesting that FD amplitude variations are influenced by both energy spectrum and geomagnetic conditions.
We suggest that low-rigidity particles, more sensitive to magnetic disturbances, show a sharp FD drop, while high-rigidity ones, less affected, exhibit a slower decline.
\section{Conclusion and summary}\label{conc}
In this study, we present the relationship between the CR intensity decreases and the associated strong geomagnetic disturbances as well as their relations with the IP parameters. For this purpose, we have used 57  minute time bins of geomagnetic index (SYM-H), FD (amplitude (\%) \& count rate) as well as various IP parameters.
High-resolution minute time bins were used to precisely calculate the time-lag/lead between the commencement of FDs and magnetic storms.
As a result, the majority of the events in our study had a time-lead (FDs leads magnetic storms) of 90 to 240 minutes, which is crucial for space weather forecasting.\\ 
We have classified the data into four groups based on the main phase decrease steps and one more different group which is CIR-driven storms.
The classified groups underwent superposed epoch analysis, which revealed that for CME-driven events, a fast, turbulent, high-field sheath structure emerged prior to the commencement of the FD and passed through during its onset.
In contrast, CIR-driven events exhibited delayed amplification, leading to significantly disturbed dynamics.\\
We classified the IP parameters into three groups: single parameters, two-parameter derivatives, and three-parameter derivatives. We performed correlation analysis between FD amplitude and the peak values of various IP parameters. As a result, FD amplitude is best predicted by the single parameter of solar wind speed (v), the two parameter derivative of IP electric field related functions (B$\times$v) and three-parameter derivative of the ratio of IMF perturbation to total magnetic field, merged with the IP electric field function [($\sigma$B/B)$\times$B$\times$v], for the three groups, respectively. Our result is consistent with previous studies \citep[e.g.,][]{2020ApJ...896..133L}.\\
For CME induced events, the relationship between FD amplitude with SYM-H minimum and amplitude have been established. The findings suggest a better correlation for moderate and strong magnetic storms compared to extreme CME-driven storms.
The manifestation of ICME (transient speed, transient period and angular size) from the ejection point to the arrival of Earth's orbit have been correlated with the FD amplitude for selected events. We found that, ICME transit velocity exhibits a stronger correlation with the FD amplitude, and ICME with fast forward shock ahead of the compression sheath region shows strong correlation than ICME without shock. \\
The FD amplitude and geomagnetic cutoff rigidity as well as energy spectra was correlated for the selected nine events for various neutron monitor stations located with different latitude and altitudes. 
\cite{2013hell.confQ..20M} reported that, the change in cutoff rigidity during FDs varies greatly depending on the neutron monitor's location. Cutoff rigidity and energy exhibit a two-step linear dependence on the amplitude of FD.
In the lower rigidity range (0.3--7.5 GV) and energy spectrum (10.17--18.5 GeV), FD amplitude decreases sharply, while in higher rigidity regimes, the decrease is more gradual. 
In addition to energy dependence, our finding confirms that, the amplitude increases linearly as the latitude of neutron monitors increase.
\backmatter
\bmhead{Acknowledgements}
We used solar wind data from the GSFC/SPDF OMNIWeb interface, available at \url{https://omniweb.gsfc.nasa.gov}. We acknowledge the world neutron monitor network for providing CR intensity data through the portal \url{https://www.nmdb.eu/nest/}. ICME characteristics were obtained from the Richardson and Cane ICME list (\url{https://izw1.caltech.edu/ACE/ASC/DATA/level3/icmetable2.htm}), SOHO/LASCO \url{https://cdaw.gsfc.nasa.gov/CME_list/} and \url{https://space.ustc.edu.cn/dreams/wind_icmes/index.php}. We thank the reviewers for their valuable feedback and constructive suggestions, which significantly enhanced the quality of this work.

\section*{Declarations}
The authors declare that there are no conflicts of interest relevant to this work.\\
\bmhead{Ethics declaration:}  not applicable.
\bmhead{Funding information} The authors declare that no funds, grants, or other support were received during the preparation of this manuscript.

\begin{appendices}

\section{Appendix}\label{A2}

For completeness of the superposed plots of all five groups, here we have displayed the rest three groups of f$^2$, f$^3$ and f$^4$ and Table \ref{t1}, as a continuous of figures from Section \ref{sub:sea}.
\begin{figure}[t]
	\centering
	\includegraphics[width=0.99\linewidth, height=9.4cm]{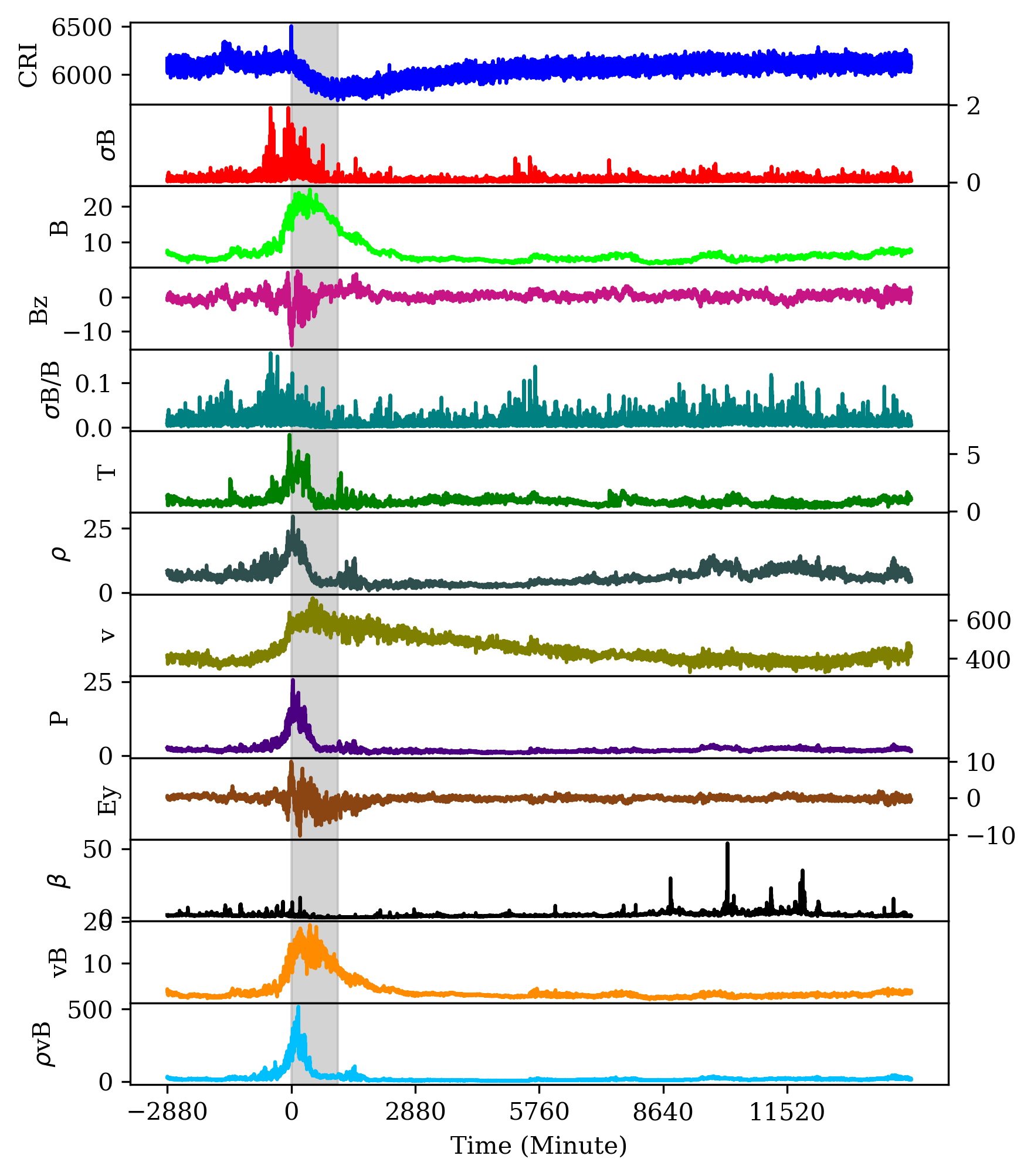}
	\caption{Similar to Figure \ref{fig:3}, except for group f$^2$.} \label{fig:4}
\end{figure}

\begin{figure}[t]
	\centering
	\includegraphics[width=0.99\linewidth, height=9.4cm]{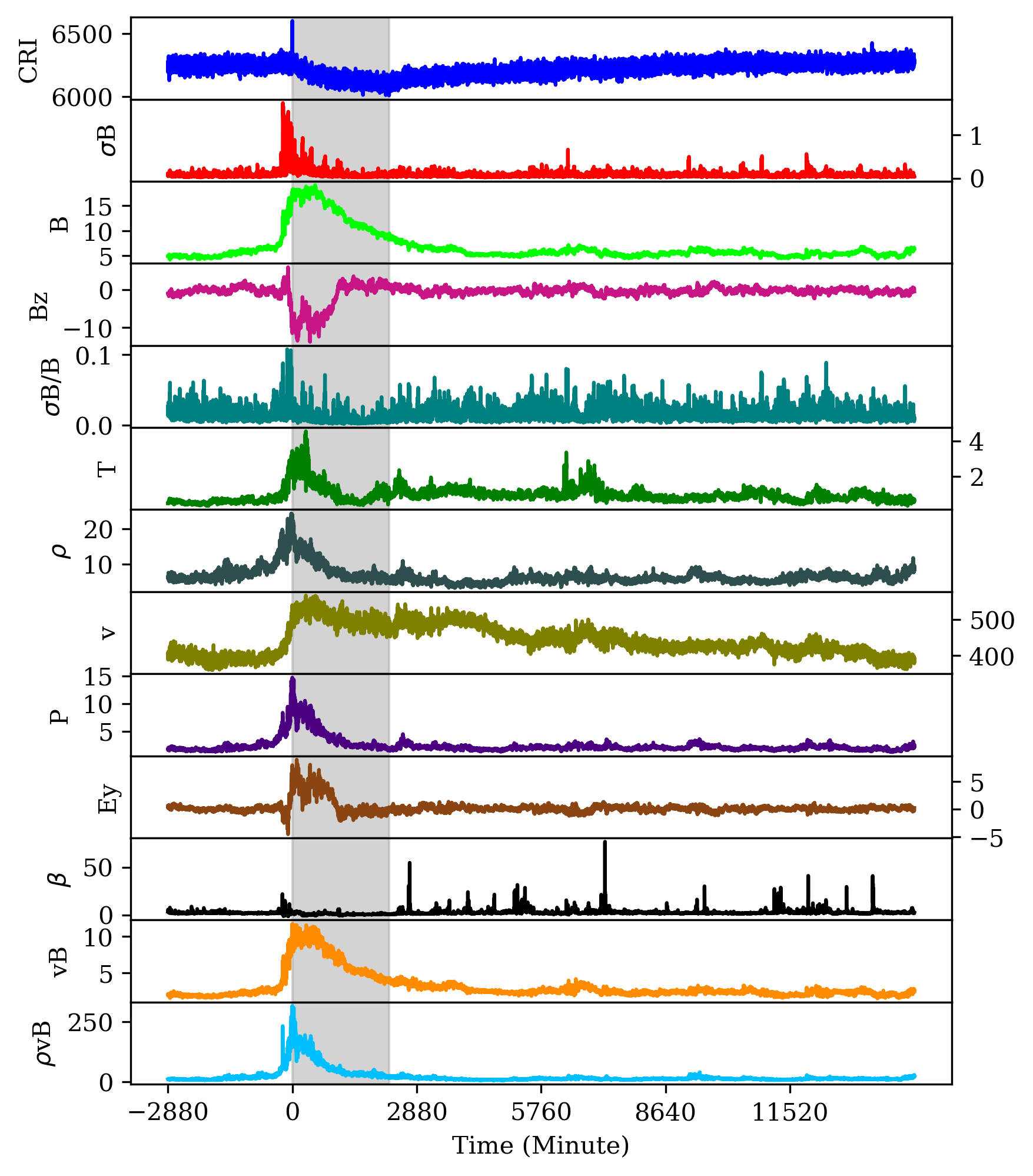}
	\caption{Similar to Figure \ref{fig:3}, except for group f$^3$.} \label{fig:5}
\end{figure}

\begin{figure}[t]
	\centering
	\includegraphics[width=0.99\linewidth, height=9.4cm]{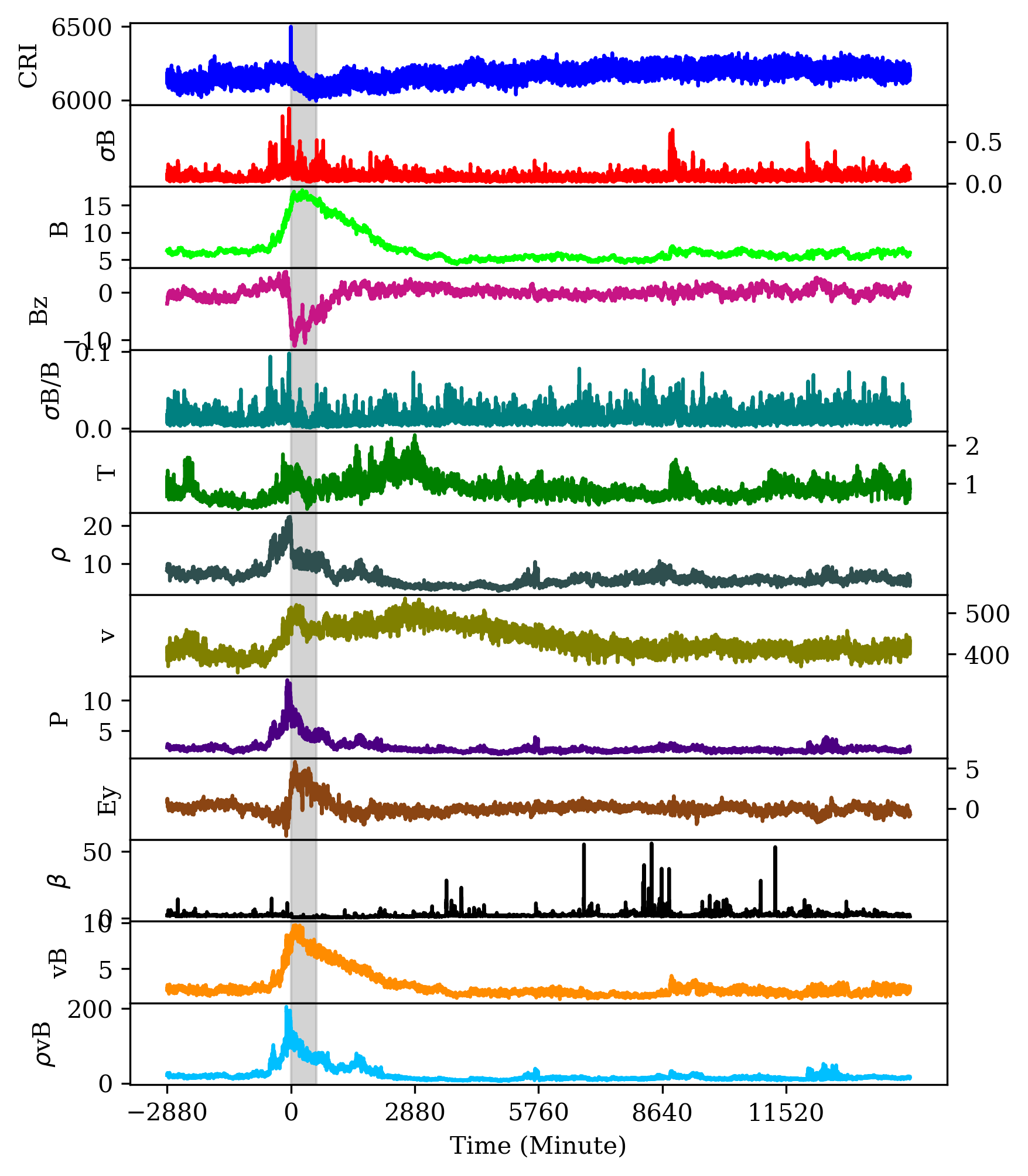}
	\caption{Similar to Figure \ref{fig:3}, except for group f$^4$.}
	\label{fig:6}
\end{figure}
\begin{table*}[t]
	\centering
	\caption{Intensity (SYM-H$_\mathrm{min}$ (nT) and Amplitude ($\Delta$SYM-H (nT) of geomagnetic storm
		and Amplitude of ($\Delta$FD (\%)) during strong geomagnetic disturbances (SYM-H $<-\bf{50}$ nT) with
		following columns.}\label{t1}
	\begin{tabular}{@{}lccccccc@{}}
		\toprule
		\multicolumn{4}{@{}c@{}}{SYM-H}&\multicolumn{3}{@{}c@{}}{FD}&Source \\
		Start date&$\Delta$SYM-H&SYM-H$_\mathrm{min}$&gp&Start Date&Ampl. (\%)&gp&\\
		\midrule
		1995-04-07T01:00 & 182  &-163& g$^3$&1995-04-07T02:00&1.9&f$^{\mathrm{CIR}}$&CIR\\
		1995-10-18T12:00 & 146  &-125& g$^3$&1995-10-18T07:00&2.2&f$^4$&CME\\
		1997-01-10T02:00 & 107 &-85&  g$^1$&1997-01-10T03:00&2.6&f$^2$&CME\\
		1997-04-21T09:00 &  92 &-100& g$^2$&1997-04-21T10:00&1.3&f$^3$&CME\\
		1997-05-01T15:00 &  90 &-80&  g$^3$&1997-05-01T13:00&1.4&f$^{\mathrm{CIR}}$&CIR\\
		1997-05-15T06:00 &  143 &-129& g$^3$&1997-05-15T07:00&1.8&f$^1$&CME\\
		1997-05-26T11:00 &  93 &-85&  g$^3$&1997-05-26T11:00&1.9&f$^3$&CME\\
		1997-06-07T01:00 &  87 &-89&  g$^4$&1997-06-07T02:00&2.4&f$^1$&CME\\
		1997-09-03T16:00 &  108 &-99&  g$^4$&1997-09-03T10:00&3.0&f$^3$&CME\\
		1997-10-10T18:00 &  126 &-139& g$^2$&1997-10-10T14:00&2.3&f$^2$&CME\\
		1997-11-06T23:00 &  118 &-125& g$^1$&1997-11-06T17:00&3.4&f$^1$&CME\\
		1998-01-06T15:00 &  89 &-84 & g$^2$&1998-01-06T11:00&3.0&f$^3$&CME\\
		1998-03-10T11:00 &  144 &-121& g$^2$&1998-03-10T12:00&2.2&f$^{\mathrm{CIR}}$&CIR\\
		1998-06-25T23:00 &  120 &-120& g$^2$&1998-06-25T13:00&1.8&f$^4$&CME\\
		1998-08-06T01:00 &  190 &-169& g$^3$&1998-08-05T19:00&2.3&f$^3$&CME\\
		1998-08-26T13:00 &  176 &-174& g$^4$&1998-08-26T11:00&7.6&f$^3$&CME\\
		1998-09-25T00:00 &  208 &-217& g$^3$&1998-09-24T19:00&10&f$^3$&CME\\
		1998-11-13T01:00 &  121 &-124& g$^4$&1998-11-13T01:00&3.2&f$^3$&CME\\
		1999-01-13T02:00 &  118 &-111& g$^4$&1999-01-12T18:00&3.1&f$^4$&CME\\
		1999-09-22T20:00 &  185 &-166& g$^2$&1999-09-22T16:00&2.6&f$^4$&CME\\
		2000-01-11T08:00 &  93  &-83 & g$^2$&2000-01-11T10:00&2.4&f$^{\mathrm{CIR}}$&CIR\\
		2000-04-06T18:00 &  305 &-320& g$^1$&2000-04-06T12:00&3.8&f$^3$&CME\\
		2000-07-15T19:00 &  351 &-347& g$^3$&2000-07-15T12:00&12.3&f$^2$&CME\\
		2000-08-12T03:00 &  223 &-235& g$^1$&2000-08-12T04:00&3.6&f$^4$&CME\\
		2000-09-17T20:00 &  199 &-203& g$^1$&2000-09-17T13:00&8.0&f$^2$&CME\\
		2001-03-31T04:00 &  452 &-437& g$^1$&2001-03-30T23:00&6.5&f$^3$&CME\\
		2001-04-11T16:00 &  270 &-280& g$^1$&2001-04-11T17:00&12.3&f$^3$&CME\\
		2001-04-18T01:00 &  122 &-122& g$^1$&2001-04-18T01:00&3.7&f$^4$&CME\\
		2001-04-22T00:00 &  114 &-104& g$^3$&2001-04-21T14:00&3.0&f$^4$&CME\\
		2001-08-17T17:00 &  140 &-131& g$^1$&2001-08-17T16:00&7.1&f$^1$&CME\\
		2001-10-03T07:00 &  134 &-188& g$^1$&2001-10-03T07:00&2.6&f$^4$&CME\\
		2001-11-05T19:00 &  326 &-320& g$^3$&2001-11-05T19:00&12.5&f$^2$&CME\\
		2001-11-24T07:00 &  223 &-234& g$^3$&2001-11-24T06:00&10.7&f$^2$&CME\\
		2002-09-07T13:00 &  159 &-168& g$^2$&2002-09-07T14:00&3.8&f$^3$&CME\\
		2003-11-20T03:00 &  488 &-490& g$^4$&2003-11-20T04:00&5.1&f$^1$&CME\\
		2004-04-03T15:00 &  147 &-149& g$^2$&2004-04-03T11:00&3.4&f$^3$&CME\\
		2004-08-30T01:00 &  143 &-128& g$^4$&2004-08-30T03:00&2.3&f$^4$&CME\\
		2005-01-21T20:00 &  82  &-101 & g$^4$&2005-01-21T18:00&8.6&f$^4$&CME\\
		2005-05-15T06:00 &  204 &-305& g$^1$&2005-05-15T02:00&10.7&f$^1$&CME\\
		2006-12-14T21:00 &  203 &-211& g$^4$&2006-12-14T15:00&9.1&f$^1$&CME\\
		2009-07-21T22:00 &  105 &-95 & g$^2$&2009-07-21T23:00&1.8&f$^4$&CME\\
		2011-08-05T20:00 &  136 &-126& g$^2$&2011-08-05T17:00&5.5&f$^2$&CME\\
		2011-10-24T22:00 &  173 &-160& g$^2$&2011-10-24T16:00&6.4&f$^1$&CME\\
		2012-04-23T18:00 &  135 &-125& g$^4$&2012-04-23T18:00&3.3&f$^4$&CME\\
		2012-07-15T01:00 &  136 &-123& g$^4$&2012-07-14T18:00&7.0&f$^2$&CME\\
		2012-11-12T19:00 &  128 &-118& g$^4$&2012-11-12T20:00&4.9&f$^4$&CME\\
		2013-03-17T07:00 &  140 &-132& g$^4$&2013-03-16T03:00&7.8&f$^4$&CME\\
		\botrule
	\end{tabular}
\end{table*}
\begin{table*}[t]
	\centering
	\begin{tabular}{@{}lccccccc@{}}
		\toprule
		\multicolumn{4}{@{}c@{}}{SYM-H}&\multicolumn{3}{@{}c@{}}{FD}&Source \\
		Start date&$\Delta$SYM-H&SYM-H$_\mathrm{min}$&gp&Start Date&Ampl. (\%)&gp&\\
		\midrule
		2013-06-01T01:00 &  141 &-137& g$^2$&2013-05-31T21:00&3.0&f$^{\mathrm{CIR}}$&CIR\\
		2013-06-06T15:00 &  82  &-88 & g$^4$&2013-06-06T12:00&3.2&f$^3$&CME\\
		2014-02-27T18:00 &  100 &-101 & g$^3$&2014-02-27T19:00&4.0&f$^{\mathrm{CIR}}$ &CIR\\
		2015-01-07T08:00 &  120 &-135& g$^1$&2015-01-07T08:00&3.2&f$^4$&CME\\
		2015-03-17T07:00 &  248 &-234& g$^3$&2015-03-17T00:00&5.7&f$^3$&CME\\
		2015-06-22T06:00 &  215 &-208& g$^4$&2015-06-22T03:00&10.8&f$^3$&CME\\
		2015-12-31T11:00 &  127 &-117& g$^4$&2015-12-31T10:00&6.1&f$^2$&CME\\
		2016-10-13T02:00 &  113 &-114& g$^1$&2016-10-12T23:00&3.5&f$^4$&CME\\
		2017-05-27T23:00 &  146 &-142& g$^1$&2017-05-27T19:00&3.1&f$^3$&CME\\
		2018-08-25T13:00 &  211 &-206& g$^4$&2018-08-25T08:00&2.6&f$^4$&CME\\
		\botrule
	\end{tabular}
\end{table*}




\end{appendices}
 \bibliographystyle{sn-mathphys-ay}
\bibliography{mple}

\end{document}